\documentclass[11pt]{article}

\usepackage{amsmath, amssymb, amsthm, amsfonts}
\usepackage{hyperref}
\usepackage{geometry}
\usepackage{authblk}
\usepackage{float}
\usepackage{graphicx,subfigure}
\usepackage{soul}

\usepackage{xcolor}

\theoremstyle{plain}

\DeclareMathOperator{\tr}{Tr}
\DeclareMathOperator{\ex}{\mathbb{E}}

\title{Bootstrapping Noncommutative Geometry with Dirac Ensembles}
\author[1]{Masoud Khalkhali}
\author[2]{Nathan Pagliaroli\footnote{\emph{Email addresses}:  masoud@uwo.ca, npagliar@uwaterloo.ca}}
\affil[1]{Department of Mathematics, The University of Western Ontario, London, ON, Canada}
\affil[2]{Department of Combinatorics and Optimization, The University of Waterloo, Kitchener, ON, Canada}
\date{}

\begin{document}
\maketitle

\begin{abstract}
This paper  surveys a bootstrap framework for random Dirac operators arising from finite spectral triples in noncommutative geometry. Motivated by a toy model for quantum gravity  to replace integration over metrics by integration over Dirac operators, we give an overview of multitrace and multimatrix random matrix models built from  spectral triples  and analyze them in the large $N$ limit using positivity constraints on  Hankel moment matrices. In this setting, the bootstrap philosophy, originating in the S-matrix program and revived in modern conformal bootstrap theory, reappears as a rigorous  analytic tool for extracting spectral data from consistency alone, without solving the model explicitly.
 We explain how Schwinger-Dyson equations, factorization at large $N$, and the noncommutative moment problem lead to finite-dimensional semidefinite programs whose feasible regions encode the allowed pairs of coupling constants and moments. Connections with  spectral geometry, in particular the study of Laplace eigenvalues,  are also discussed, illustrating how bootstrapping provides a unified mechanism for deriving bounds in both commutative and noncommutative settings.

\end{abstract}

\medskip

\tableofcontents

\section{Introduction}

The past two decades have witnessed a  remarkable  connection  between high-energy theoretical physics, noncommutative  geometry, and random matrix theory. At the center of this convergence is a profound insight of Connes: in noncommutative geometry, the Dirac operator encapsulates the fundamental geometric data of a space. This idea reaches its full expression in the {\it spectral action principle}, in which the metric, gauge fields, and Higgs degrees of freedom arise from the spectral properties of the Dirac operator \cite{Connes94, Connes95, NCG and Reality, Connes Inner Fluctuations, ConnesRecon, Spectral action, Neutrino Mixing}. Allowing this operator to fluctuate therefore amounts to quantizing geometry itself. In parallel, the revival of bootstrap philosophy in conformal field theory has demonstrated the power of positivity, symmetry, and spectral consistency to constrain a physical theory without the need for explicit dynamical solutions \cite{RRTV, bootstraps}. These developments together suggest that fluctuating Dirac operators may be studied through a combination of spectral, geometric, and positivity-based methods, borrowing intuitions from both random matrix theory and modern bootstrap techniques.

This survey aims to present a unified view of these themes through the study of \emph{Dirac ensembles}, in which the Dirac operator plays the role of the random geometric field. First initiated by the proposal of Barrett and Glaser to integrate over finite-dimensional Dirac operators \cite{Barrett2015, Barrett2016, glaser}, we examine how ideas from multitrace random matrix theory, spectral geometry,  and moment-positivity methods  can be adapted to analyze such ensembles \cite{Azafar2024,  hessam2022bootstrapping,  hessam2022fromnoncom, hessam2023double,  bootstraps, bootstraps berenstein, bootstraps Kazakov, bootstraps critical behaviour}.

 In their study, Barrett–Glaser  used  Monte-Carlo Markov chain simulations to investigate the phase structure of their random finite spectral triples. In a complementary direction, the approach taken in our work and related papers relies on analytic and probabilistic methods from random matrix theory, such as multitrace genus expansions, Coulomb-gas techniques, topological recursion, and bootstrap constraints, offering a more explicitly analytic perspective on the geometry of Dirac ensembles for certain families of models.

The philosophy is that, just as the spectral action expresses the classical dynamics of an almost-commutative geometry \cite{Spectral action, Neutrino Mixing, VS}, a Dirac ensemble defines a non-perturbative measure on the space of geometries. In this approach the task is not to solve the 
model exactly,  which is generally impossible for multitrace multimatrix ensembles, but rather to delimit the space of allowable moment data using positivity inherited from the Hilbert space representation and consistency inherited from the Schwinger--Dyson equations.

The analogy with bootstrap methods is not superficial. In spectral geometry, positivity of integrated eigenfunction correlations can be used to bound Laplace eigenvalues on surfaces and 3-manifolds \cite{Bon1, Bon2}. In conformal field theory, positivity of squared OPE coefficients constrains the allowed spectra of primary operators \cite{RRTV}. In random matrix theory, Hankel-matrix positivity restricts the eigenvalue distributions compatible with a given potential. Dirac ensembles provide a meeting ground for these ideas: the Schwinger--Dyson relations play the role of crossing symmetry, while moment positivity arises from the structure of finite noncommutative geometries \cite{hessam2022bootstrapping, hessam2022fromnoncom}. Together these yield a ``spectral bootstrap’’ for noncommutative geometry.

Another central theme of the paper is the incorporation of gauge and Higgs degrees of freedom into Dirac ensembles. While the simplest ensembles fluctuate only the metric sector of a fuzzy spectral triple \cite{Barrett2015, Barrett2016}, one can couple the fuzzy geometry to a finite spectral triple in the spirit of almost-commutative manifolds \cite{Connes-Lott, Spectral action, VS}. The result is a \emph{Yang--Mills--Higgs Dirac ensemble} in which both geometry and gauge fields fluctuate through inner perturbations of the Dirac operator \cite{Connes Inner Fluctuations, Sanchez Yang-Mills}. The corresponding spectral action resembles a discretized Yang--Mills--Higgs Lagrangian \cite{Spectral action, Sanchez Yang-Mills}, and many aspects of Standard Model geometry can be emulated in this finite setting \cite{Finite spectral triples, Neutrino Mixing}. These models are analytically challenging and provide fertile ground for applying the bootstrap and moment-positivity techniques introduced earlier.

 This exposition is intended to be accessible, situating Dirac ensembles within the broader context of spectral geometry, fuzzy manifolds, and random matrix models.  Here is a section-by-section overview of the paper.
 
Section 2 begins with a brief historical account of the bootstrap philosophy, starting from Chew’s S-matrix program \cite{Chew1962}, and continuing to modern developments in the conformal bootstrap \cite{RRTV}. This section sets the conceptual tone by illustrating how far-reaching constraints can be extracted from consistency and positivity alone.

Section 3 introduces Dirac ensembles as finite-dimensional models of fluctuating geometry. After reviewing fuzzy spectral triples and the Barrett--Glaser proposal \cite{Barrett2015, Barrett2016, glaser}, we describe the multitrace and multimatrix structures that naturally arise. 
%and discuss their large $N$ limits \cite{Khalkhali2020, Khalkhali2022, Khalkhali2024 coloured maps}.

Section 4 develops a bootstrap framework for Dirac ensembles. Schwinger--Dyson equations are combined with positivity techniques \cite{bootstraps, bootstraps berenstein} to obtain constraints on allowable moment sequences. Several explicit models are worked out to illustrate both symmetric and symmetry-breaking behaviors \cite{hessam2022bootstrapping, hessam2022fromnoncom}.

Section 5 considers the bootstrap analysis to multimatrix quantum mechanical models, drawing parallels with recent work on the bootstrap for matrix quantum mechanics and lattice gauge theories \cite{bootstraps matrix quantum mechancis cho, bootstraps Guo lattice yang-mills, LinZheng2025}.

Section 6 studies Dirac ensembles formed by coupling a fuzzy spectral triple to a finite ones \cite{Connes-Lott, Sanchez Yang-Mills}. After reviewing the finite geometry relevant to the Standard Model \cite{Neutrino Mixing, VS}, we describe the structure of inner fluctuations \cite{Connes Inner Fluctuations}, the emergence of Higgs-type fields \cite{Spectral action, Sanchez Yang-Mills}, and the coupling to fermions \cite{Barrett2024, Khalkhali2025 large N limit}.

Section 7 turns to spectral geometry and explains how bootstrap ideas manifest in the analysis of the eigenvalues Laplacian. By examining positivity properties of integrated eigenfunction correlations  one can constrain spectral gaps on surfaces and three-dimensional manifolds \cite{Bon1, Bon2}.

Section 8 concludes with open problems, conjectures, and future directions.

\section{Origins of the Bootstrap Program}
\label{sec:bootstrap-origins}

The word \emph{bootstrap} first entered   theoretical physics in the early 1960s, in the context of Geoffrey Chew's S-matrix program for describing the strong nuclear force \cite{Chew1962}. The central idea was radical for its time: rather than postulating a fundamental Lagrangian or a set of elementary fields, one should attempt to determine all observable quantities directly from general consistency principles such as unitarity, analyticity, and crossing symmetry of scattering amplitudes. Chew famously summarized this view by saying that ``there are no elementary particles'' and that the hadronic spectrum should be determined self-consistently, ``particles pull themselves up by their own bootstraps.''

The S-matrix bootstrap achieved partial success: it correctly organized the hadronic resonance spectrum, led to Regge theory, and indirectly pointed the way to the discovery of the Veneziano amplitude, the first appearance of what later became string theory. But the program ultimately failed to produce a unique or predictive description of strong interactions. With the discovery of asymptotic freedom and the rise of quantum chromodynamics in the 1970s, the bootstrap ideology retreated from particle physics.

A major revival occurred in 1983, when Belavin, Polyakov, and Zamolodchikov (BPZ) introduced the \emph{conformal bootstrap} in two dimensions \cite{BPZ1984}. There the space of local fields in a conformal field theory (CFT) is constrained by the Virasoro algebra, and the consistency of operator product expansions leads to exact non-perturbative solutions. For nearly twenty years the method remained confined to two dimensions, until the landmark work of Rattazzi, Rychkov, Tonni, and Vichi in 2008 \cite{RRTV}, who demonstrated that a variant of the bootstrap based on convexity, positivity of operator coefficients, and numerical semidefinite programming could be used to extract rigorous bounds on operator dimensions in higher-dimensional CFTs. This initiated what is now called the \emph{numerical bootstrap program}, a field that has rapidly developed into a central tool in modern conformal field theory and has inspired related bootstrap approaches in random matrix models, lattice gauge theories, matrix quantum mechanics and more \cite{bootstraps,bootstraps Kazakov,bootstraps matrix quantum mechancis cho,bootstraps Li lattice gauge,bootstraps Guo lattice yang-mills,bootstraps Kazakov finite lattice,boostraps lattice yang mills,bootstraps Li analytic trajectory,bootstraps Li Yang-Mills matrix integrals}.

A bootstrap problem can be expressed schematically as the task of finding a set of unknown spectral quantities $\{c_\alpha\}$ satisfying a consistency equation
\[
\sum_{\alpha} c_\alpha\, F_\alpha(x) = 0,
\]
together with positivity conditions $c_\alpha \ge 0$ coming from unitarity, measure positivity, or geometric constraints. The underlying question is then: \emph{is the system solvable, and if so, is the solution unique?} If no solution exists under the assumed spectral ansatz, the corresponding theory is ruled out. If a solution exists, the theory is \emph{bootstrapped} into existence.

This abstract formulation provides the conceptual bridge to the mathematical applications developed later in this paper. The central principle shared across these developments is that \emph{positivity} and \emph{consistency} constraints, when combined with symmetry, may be strong enough to determine or severely restrict the underlying theory. Just as the conformal bootstrap constrains operator dimensions \cite{BPZ1984,RRTV}, one may constrain the eigenvalue spectrum of random matrices or the Laplacian eigenvalues of negatively curved manifolds, or heat coefficients \cite{bootstraps,bootstraps berenstein,Bon1,Bon2,hessam2022bootstrapping,Khalkhali2022}.

% ================================

\section{Dirac Ensembles}
\subsection{Path Integral Quantization and the Role of Noncommutative Geometry}
\label{sec:path-integral-ncg}

In the standard formulation of quantum field theory, the expectation value of a classical observable
$\mathcal{O}$ is defined by the Feynman path integral
\[
\langle \mathcal{O} \rangle \;=\; \int D[\varphi] \, \mathcal{O}(\varphi)\, 
\exp\!\left( \frac{i}{\hbar} S(\varphi)\right),
\]
where the integral ranges over all field configurations $\varphi$ and $S(\varphi)$ denotes the classical action functional. 
Although this expression is formally ill-defined as written, it provides the conceptual foundation for perturbative
quantization, Wick rotation, and Euclidean functional integrals.

In gauge theories such as Yang--Mills, the divergence of the functional integral can be controlled by 
renormalization: after introducing a regularization scheme  such as a momentum cutoff, Pauli--Villars regulator, heat kernel
cutoff, dimensional regularization, etc. where the ultraviolet divergences can be absorbed into a finite number of
counterterms. This is the reason that quantum electrodynamics and the Yang--Mills sector of the Standard Model
are mathematically consistent quantum field theories: the renormalized action contains only finitely many
independent parameters, and the perturbation series remains predictive.

The situation is quite different for gravity. If one attempts to quantize the metric field $g_{\mu\nu}$ using 
the Einstein--Hilbert action
\[
S(g) = \int_M R(g)\, \mathrm{dvol}_g,
\]
one finds that the resulting quantum field theory is non-renormalizable. That is, at each order of perturbation theory,
new counterterms of increasingly high dimension appear, and no finite parametrization of the theory exists.
From the viewpoint of perturbative quantum field theory, this is the fundamental obstruction to a
 quantization of classical general relativity.

This difficulty motivates the search for alternative formulations of quantum gravity in which the action is 
regularized at small scales or replaced by a more geometric object with better ultraviolet behavior. One such
proposal, developed by Connes and collaborators, is the \emph{spectral action principle} in noncommutative
geometry \cite{Spectral action, Neutrino Mixing, ConnesRecon}, in which the classical Einstein--Hilbert action is replaced by the trace of a function of the Dirac
operator on a noncommutative space. In this framework, the analog of the path integral is not a functional
integral over metrics but a spectral partition function depending on the eigenvalues of a generalized Dirac
operator. The ultraviolet divergences are then controlled by heat kernel asymptotics rather than perturbative
counterterms, and the Standard Model of particle physics emerges from the geometry of an ``almost commutative''
space \cite{Connes-Lott, QFTNCG, VS}.

% --- INSERTED PARAGRAPH ---
\medskip
In the finite-dimensional setting of noncommutative geometry, this analogy leads naturally to the notion of a \emph{Dirac ensemble}, where the role of the field configuration is played by a fluctuating finite Dirac operator. This idea was first explored numerically in the Monte Carlo studies of Barrett and Glaser \cite{Barrett2016, glaser, truncated 1, truncated 2}, and later developed analytically through random matrix and probabilistic techniques such as Coulomb-gas methods and topological recursion \cite{Khalkhali2020, Khalkhali2022, hessam2023double,AKHZ}. In this framework the path integral becomes a matrix integral over Dirac operators, with geometry encoded in their 
spectra. We should also mention that  within the Wulkenhaar-Grosse program for defining quantum field theory on noncommutative spaces, the formalism inherently involves random matrices in the analysis of the effective 
action. We refer to \cite{Wulkenhaar} and references therein for more details. It is an interesting problem to find a bridge between the Dirac ensemble approach and the noncommutative quantum field theory. 

\medskip
% --- END INSERTED PARAGRAPH ---

The remainder of the paper will return to this analogy: just as renormalization corresponds to subtracting
divergent terms in a heat kernel expansion, the bootstrap method of Sections \ref{sec:bootstrap-dirac} and
\ref{sec:eigenfunction-bootstrap} may be viewed as a way of constraining spectral data without reference to
a Lagrangian. In both cases, the basic objects are not fields but eigenvalues, and the central constraints arise
from positivity.

% ================================

\subsection{Metric from Dirac Operators and the Spectral Formulation of Geometry}
\label{sec:dirac-metric}

One of the central insights of noncommutative geometry is that the Riemannian metric of a spin
manifold can be completely recovered from the spectral properties of its Dirac operator.  Let
$(M,g)$ be a compact Riemannian spin manifold, let $D$ denote the Dirac operator acting on the
Hilbert space $\mathcal{H}=L^{2}(M,S)$ of square-integrable spinors, and let $C^{\infty}(M)$ act on
$\mathcal{H}$ by pointwise multiplication.  Connes' distance formula expresses the geodesic distance
between points $p,q\in M$ as
\[
d(p,q)
  = \sup\left\{
      |f(p)-f(q)| : f\in C^{\infty}(M),\ \|[D,f]\|\le 1
    \right\}.
\]
Since the commutator $[D,f]$ is Clifford multiplication by $\nabla f$, the condition
$\|[D,f]\|\le 1$ is equivalent to $|\nabla f|\le 1$.  Thus the spectral expression reproduces the
classical Kantorovich–Rubinstein dual formulation of geodesic distance.  In this sense, the Dirac
operator encodes not only the Laplace spectrum but the full Riemannian metric i.e. distance, volume,
dimension, and eventually curvature can all be recovered from $D$.  This  reformulates the Riemannian distance in a path-free manner by writing in terms of the Lipschitz bound
\[
\|[D,f]\| \le 1 \quad \Longleftrightarrow \quad |\nabla f| \le 1,
\]
so the right-hand side is the same as the classical dual formulation of the geodesic distance:
\[
d(p,q)
  \;=\;
  \inf_{\gamma} \int_\gamma \sqrt{g_{\mu\nu}\, dx^\mu dx^\nu},
\]
where the infimum ranges over all smooth paths $\gamma$ from $p$ to $q$. In this sense, the Dirac operator
encodes all information about the underlying metric, not only the spectrum of the Laplace operator but the
actual point-to-point distance.

This leads to an operator-theoretic reformulation of Riemannian geometry.  The triple $(A,\mathcal{H},D)$
where $A$ is a $*$-algebra of ``coordinates,'' $\mathcal{H}$ a Hilbert space carrying a
representation of $A$, and $D$ a self-adjoint operator with bounded commutators $[D,a]$ and compact
resolvent, is called a \emph{spectral triple}.  When $A=C^{\infty}(M)$, one recovers ordinary spin
geometry.  When $A$ is noncommutative, one obtains geometries with noncommuting coordinates. Spectral triples were introduced by Connes in \cite{NCG and Reality} (see also
\cite{QFTNCG}) as the fundamental objects of noncommutative Riemannian geometry.

A fundamental result is Connes’ \emph{Reconstruction Theorem}, proved in \cite{ConnesRecon}. To fully reconstruct a Riemannian spin manifold from a spectral triple, some additional structure is required. 
A \emph{real spectral triple} is a spectral triple equipped with two additional operators: an anti-unitary map  $J:\mathcal{H}\to\mathcal{H}$, called the \textit{real structure} and a self-adjoint operator $\gamma$ on $\mathcal{H}$ with $\gamma^2=1$, called the \emph{chirality operator} or
\emph{grading}. The chirality operator and real structure must satisfy the order-zero condition  
\[
[a,\, J b J^{-1}] = 0, \qquad \forall\, a,b\in\mathcal{A},
\]
and the order-one condition
\begin{equation*}
	[[D,a],J^{-1}b^{*}J]=0
\end{equation*}
for all $a,b \in \mathcal{A}$. We further require that: 
\begin{itemize}
	\item $J^2 = \varepsilon$
	\item $JD = \varepsilon' DJ$
	\item $J\gamma = \varepsilon'' \gamma J$
\end{itemize}
where the choice of signs $\varepsilon,\varepsilon'$, and $\varepsilon''$ is dependent on the KO dimension via Table \ref{tab:KO_dimension} in the even case and in the odd case they are trivial. 
\begin{table}[h]
	\centering
	\begin{tabular}{r | c c c c c c c c}
		$s \mod 8$ 		& 0 & 1 & 2 & 3 & 4 & 5 & 6 & 7 \\
		\hline
		$\epsilon$		& $+$ & $+$ & $-$ & $-$ & $-$ & $-$ & $+$ & $+$ \\
		$\epsilon'$		& $+$ & $-$ & $+$ & $+$ & $+$ & $-$ & $+$ & $+$ \\
		$\epsilon''$	& $+$ & $+$ & $-$ & $+$ & $+$ & $+$ & $-$ & $+$
	\end{tabular}
	\caption{The standard choices of the signs associated to a $KO$-dimension  \cite{Connes95}.}. 
	\label{tab:KO_dimension}
\end{table}

These conditions generalize the relations of the same objects in the case of spin$^c$ manifolds. More generally, real spectral triple can be viewed as a noncommutative analogue of a
compact spin$^c$ Riemannian manifold. Connes' reconstruction theorem  asserts that, conversely, any
commutative real spectral triple satisfying mild regularity conditions is uniquely
equivalent to the spectral triple of a spin$^c$ manifold.  Thus, spectral triples
extend the notion of Riemannian geometry to genuinely noncommutative spaces.

% In particular, when $M$ is such a manifold, then $\mathcal{A}=C^\infty(M)$, $
%\mathcal{H}=L^2(M,S)$, and $D$ is the dirac operator on spinors,
%with $J$ and $\gamma$ the standard charge conjugation and chirality operators. 
\medskip 

\noindent {\bf Theorem:} {\it Every commutative, real, $p$-summable, regular, oriented spectral triple satisfying the first-order
condition, finiteness, absolute continuity, and Poincaré duality is isomorphic to the canonical
spectral triple of a compact oriented smooth Riemannian spin manifold.  Thus the data
$(C^{\infty}(M),L^{2}(M,S),D)$ is not only an example but the \emph{unique} commutative solution to
the axioms of noncommutative geometry.}

\medskip

This provides the conceptual justification for treating spectral triples as generalized noncommutative
manifolds: the axioms recover classical spin geometry exactly when the algebra is commutative.

\bigskip
\noindent\textbf{Example 1: The Two-Point Space }

\medskip
The simplest noncommutative geometry is the discrete space $X=\{p,q\}$ with algebra
$A=\mathbb{C}^{2}$.  The Hilbert space is $\mathbb{C}^{2}$ and the algebra acts diagonally.  A Dirac
operator is chosen as
\[
D
  =
  \begin{pmatrix}
  0 & m \\
  m & 0
  \end{pmatrix},
  \qquad m>0.
\]
For $f=(f_{p},f_{q})$, the commutator is
\[
[D,f]
  =
  \begin{pmatrix}
  0 & m(f_{p}-f_{q}) \\
  -m(f_{p}-f_{q}) & 0
  \end{pmatrix},
\]
with $\|[D,f]\| = m |f_{p}-f_{q}|$. The condition $\|[D,f]\|\le 1$ becomes $|f_{p}-f_{q}| \le 1/m$.  Connes' distance formula therefore
gives
\[
d(p,q)=\frac{1}{m}.
\]
Thus the single parameter $m$ encodes the metric on a two-point space.  In the Connes--Lott model of
the Standard Model, this finite geometry becomes the ``internal space,'' with $m$ replaced by matrices
related to the Yukawa couplings.  The Higgs field then emerges naturally from inner fluctuations of
the metric, demonstrating how algebraic data can produce physically measurable quantities.

\bigskip
\noindent\textbf{Example 2: The Noncommutative Two-Torus.}

\medskip
The noncommutative 2-torus $\mathbb{T}^{2}_{\theta}$ is generated by two unitaries $U,V$ with
\[
VU = e^{2\pi i \theta} U V, \qquad \theta\in\mathbb{R}\setminus\mathbb{Q}.
\]
The smooth algebra $A_{\theta}$ consists of rapidly decaying Fourier series
\[
a = \sum_{m,n} a_{mn} U^{m} V^{n}.
\]
Derivations play the role of coordinate vector fields:
\[
\delta_{1}(U)=2\pi i U,\quad \delta_{1}(V)=0,
\qquad
\delta_{2}(U)=0,\quad \delta_{2}(V)=2\pi i V.
\]
The Dirac operator is modeled on the flat Dirac operator:
\[
D = \sigma^{1}\delta_{1} + \sigma^{2}\delta_{2}.
\]
The triple $(A_{\theta},\mathcal{H},D)$ is a genuine noncommutative spin geometry.  It reproduces the
same metric dimension and heat-kernel asymptotics as the ordinary torus but yields a genuinely
quantum metric on the state space when $\theta$ is irrational.  This spectral triple also appears in the
Connes--Douglas--Schwarz matrix model, encoding T-duality and the geometry of D-branes in a
nontrivial $B$-field background.

\medskip
The examples above illustrate that once geometry is encoded spectrally, the Dirac operator becomes
a natural dynamical variable.  In classical geometry the metric $g$ varies in the Einstein--Hilbert
functional; in the spectral formulation, $D$ plays this role.  In both the spectral action principle and
the Dirac ensemble framework, one integrates over families of Dirac operators, treating geometry (or
micro-geometry) as a fluctuating quantity.  This observation provides a conceptual bridge between
noncommutative geometry, matrix models, and random geometry: the Dirac operator contains the
metric, curvature, and gauge structure, and so fluctuating $D$ means fluctuating geometry.

The next sections will build on these examples to develop Dirac ensembles, fuzzy spectral triples, and
spectral action functionals as tools for both Euclidean quantum gravity and noncommutative
geometric models of particle physics.

\subsection{Fuzzy Spectral Triples}

A spectral triple is said to be \emph{finite} if both $\mathcal{A}$ and $\mathcal{H}$ are
finite-dimensional.  The question of classifying  finite real spectral triples was fully answered by Krajewski in \cite{Finite spectral triples}. We are interested in a particular distinguished type of real finite spectral triples known as  fuzzy spectral triples
or fuzzy geometries, introduced by
Barrett in \cite{Barrett2015}. Morally, fuzzy geometries should be thought of as
finite-resolution approximations to classical spin$^c$ manifolds. The
algebra of functions is replaced by a matrix algebra and the Dirac
operator is replaced by a finite dimensional matrix.

Let us now recall the explicit algebraic form of fuzzy spectral triples. Consider the real Clifford algebra $\mathrm{C\ell}_{p,q}$ associated to
$\mathbb{R}^{p+q}$ equipped with a quadratic form of signature $(p,q)$. Its complexification $\mathbb{C}\ell_n := \mathrm{C\ell}_{p,q} \otimes_{\mathbb{R}} \mathbb{C}$ where $n =p+q$, has the standard basis $\{ e_i \}_{i=1}^{n}$. 
Let $\gamma \in\mathbb{C}\ell_n$ be the chirality element defined by
\[
\Gamma = i^{\frac{1}{2}s(s+1)}\, e_1 e_2 \cdots e_n ,\qquad
s \equiv q - p \pmod 8.
\]
The integer $s$ is called the KO-dimension.  Consider the unique irreducible Hermitian $\mathrm{C\ell}_{p,q}$-module $V_{p,q}$.  When $n=p+q$ is odd, $\gamma$ acts on it trivially.  Additionally, the Clifford module $V_{p,q}$  has a natural charge conjugation operator
$C : V_{p,q}\to V_{p,q}$.

 A \textit{fuzzy geometry} of signature $(p, q)$  is a real spectral triple of $K O$-dimension $s \equiv q - p \pmod 8$ of the form
$$
\left(M_N(\mathbb{C}), M_N(\mathbb{C}) \otimes V_{p,q}, D ; 1 \otimes J_V\right)
$$
where $J_V$ is the real structure of $V_{p,q}$. When $p+q$ is even, the fuzzy geometry has the grading $\Gamma=1 \otimes \Gamma_V$. Many so called ``fuzzy spaces", such as the fuzzy sphere \cite{fuzzy sphere, fuzzy sphere Torres} or fuzzy torus \cite{Fuzzy torus generalized}, can be realized as fuzzy geometries \cite{Fuzzy torus as spectral triple,Barrett2015,glaser deform}.

\subsection{Dirac Ensembles}
In \cite{Barrett2015},  Barrett and Glaser study path integrals over fuzzy geometries as toy models of Euclidean quantum gravity. Their starting point is the observation, recalled in Section~\ref{sec:dirac-metric}, that the Dirac operator plays the role of the metric in a spectral triple. If geometry is to be quantized, it is therefore natural to treat the Dirac operator as the fundamental dynamical variable, and to define the gravitational partition function by
\[
Z \;=\; \int_{\text{metrics}} e^{-S(g)}\, \mathcal{D}g
\qquad\Longrightarrow\qquad
Z \;=\; \int_{\text{Dirac operators}} e^{-\mathcal{S}(D)}\, dD.
\]
In the Barrett--Glaser formulation, this idea is implemented not on an infinite-dimensional space of Dirac operators on a manifold, but on the space of Dirac operators associated to a \emph{finite real spectral triple}. As mentioned before, such triples arise naturally in the internal (noncommutative) part of the spectral model of the Standard Model, and have the form
\[
(A,\mathcal{H},D), \qquad 
A \cong \bigoplus_i M_{n_i}(\mathbb{C}), \qquad
\mathcal{H}\cong \mathbb{C}^N, \qquad
D = D^\ast \in M_N(\mathbb{C}).
\]
 Together with a grading and real structure,  they  satisfy the axioms of noncommutative spin geometry. In this context, the space of admissible Dirac operators is a finite-dimensional real vector space, so the formal measure $dD$ may be taken to be a Lebesgue measure, and the path integral becomes an honest finite-dimensional matrix integral.

More precisely, it was shown by Barrett \cite{Barrett2015}, that the Dirac operator of any fuzzy geometries must necessarily be of the form:
\begin{equation*}
D(M \otimes \psi) =\sum_{I}(K_{I}H + \epsilon'H K^{*}_{I})\otimes \left(\prod_{i\in I}\gamma_{i} \right)\psi
\end{equation*}
where the sum is over specific subsets $I \subset \{1,...,p+q\}$ depending on the signature $(p,q)$,$\gamma_{i}$ are the associated gamma matrices, and $K_{I}$ are arbitrary skew- or self-adjoint matrices. See \cite{Barrett2015} for more details. This result is foundational in establishing probability distributions over spaces of fuzzy geometries since it clearly parameterizes all the Dirac operators of fuzzy geometries as some finite sum of commutators or anti-commutators of skew- or self-adjoint matrices tensored to gamma matrices. We emphasize that the choice of skew- or self-adjoint matrices is completely arbitrary so that is free information we want to integrate over. 

To turn this into a dynamical model, one introduces an action functional $\mathcal{S}(D)$ depending on $D$ and studies the resulting probability measure
\[
\mu(D) \;=\; \frac{1}{Z}\, \exp\!\bigl(-\mathcal{S}(D)\bigr)\, dD.
\]
 A particularly simple example is the quartic spectral action
\[
\mathcal{S}(D)
  \;=\; g\, \mathrm{Tr}(D^2) \,+\, \mathrm{Tr}(D^4),
\]
where $g$ is a real coupling constant. Since $D$ is a finite self-adjoint matrix, this defines a probability distribution on a space of matrices that is reminiscent of random matrix models, but with the additional algebraic constraints imposed by the spectral triple axioms. Barrett and Glaser showed that for such models, one may study phase transitions, eigenvalue distributions, and emergent notions of geometry in a setting that lies halfway between matrix models and full noncommutative geometry. 

The resulting objects have come to be known as \emph{Dirac ensembles}: random matrix models in which the random variable is not a generic Hermitian matrix but one satisfying the structural constraints of a real spectral triple. From this point of view, the Barrett--Glaser model provides a matrix-regularized version of the spectral action for quantum gravity, in close analogy with the way ordinary Hermitian matrix models arise as discrete approximations to two-dimensional quantum gravity. 

This finite-dimensional approach has two key advantages. First, unlike the continuum Einstein--Hilbert functional integral, the partition function is mathematically well-defined. Second, the probabilistic and spectral tools of random matrix theory become available, including moment methods, large deviation theory, and bootstrap constraints of the type that will be introduced in Section \ref{sec:bootstrap-dirac}. The study of Dirac ensembles therefore lies at the intersection of noncommutative geometry, quantum gravity, and random matrix theory, and serves as a laboratory for testing ideas about the quantization of geometry itself.

% ================================
\subsection{Multitrace and Multimatrix Dirac Ensembles}
\label{sec:multitrace}

The quartic spectral action introduced in the previous section becomes considerably more
complicated when expressed in terms of the matrix variables that parametrize the space of Dirac operators.
Even in the simplest nontrivial example studied by Barrett and Glaser, the action does not reduce to a
single-trace polynomial in a single Hermitian matrix, but instead produces a combination of multitrace
terms \cite{Sanchez spectral action}. Consider A Dirac ensemble that corresponds to probability distribution over the space of all Dirac operators fuzzy geometries of type $(1,0)$. By Barrett's classification theorem, $D = \{H,\cdot\}$, where $H$ is some Hermitian matrix. We assign the probability density 
\begin{equation*}
	\frac{1}{Z}e^{-S(D)}dD
\end{equation*}
where 
\begin{equation}
\label{eq:BG-multitrace}
S(D)
  = 2N\bigl(g\, \mathrm{Tr}(H^2) + \mathrm{Tr}(H^4)\bigr)
    \;+\; 2g\, (\mathrm{Tr} H)^2
    \;+\; 8\,\mathrm{Tr}(H)\,\mathrm{Tr}(H^3)
    \;+\; 6\,(\mathrm{Tr} H^2)^2,
\end{equation}
and the measure $dD=dH$ is the Lebesgue measure on the space of $N \times N$ Hermitian matrices:
\begin{equation*}
	d D = d H = \prod_{i=1}^N d H_{ii} \, \prod_{1 \leq i < j \leq N}  d ({\text{Re}} (H_{ij})) \, d ({\text{Im}} (H_{ij})). 
\end{equation*}
Barrett and Glaser analyzed this model and other quartic models of higher type numerically using Markov Chain Monte Carlo methods and observed
phase transitions in the eigenvalue distribution as the coupling constant \(g\) varies. In general, the analytic study of such models is formidable. The difficulty arises from the fact that most classical results in random matrix theory apply only to
single-trace, single-matrix models of the form
\[
Z \;=\; \int_{\mathcal{H}_N} \exp\!\bigl( -N \mathrm{Tr} \, V(H) \bigr)\, dH,
\]
where \(V\) is a polynomial (or analytic) potential. Such models admit powerful methods such as
orthogonal polynomials, Coulomb gas methods, and topological recursion, all of which rely critically on the
existence of a single trace term. By contrast, the action \eqref{eq:BG-multitrace} contains nonlinear
interactions among traces of different powers of \(H\), and can no longer be reduced to a one-body
Coulomb gas model. In this sense, Dirac ensembles fall outside the standard universality classes of
random matrix theory.

Moreover, in more elaborate spectral triples the Dirac operator decomposes not into a single matrix
\(H\) but into several blocks, producing \emph{multimatrix models} in addition to multitrace effects.
Thus, the analytic study of Dirac ensembles requires tools capable of handling coupled matrix systems
with nonlinear trace interactions;an area in which only partial results are known for very specific single trace multimatrix models, and where the
bootstrap methods discussed later in this paper become relevant.

% ================================
\subsection{Large-\texorpdfstring{$N$}{N} Limits and Eigenvalue Distributions}

For a probability measure on matrices of the form
\[
d\mu(H) \;=\; \frac{1}{Z_N}\, e^{-S(H)}\, dH,
\]
a central object of study is the mean empirical eigenvalue distribution
\[
\rho_N(x)
  \;=\;
  \int_{\mathcal{H}_N}
    \mu_N(H)\, d\mu(H)\]
where
\[\mu_N(H) \;=\; \frac{1}{N}\sum_{i=1}^N \delta(x - \lambda_i(H)),
\]
where $\lambda_i(H)$ are the eigenvalues of $H$. In classical random matrix ensembles (Gaussian, Wigner,
Laguerre, etc.), this distribution converges as $N\to\infty$ to a deterministic limit such as the
Wigner semicircle or Marchenko–Pastur law. 

%For multitrace models such as \eqref{eq:BG-multitrace}, however, no such universal law is known, and even the existence of a limiting density must be proven from first principles.

In practice, one studies  the large-\(N\) limit through tracial moments:
\[
\lim_{N\to\infty}
\ex \left[\frac{1}{N}\mathrm{Tr}(H^k) \right],
\]
which often determine the limiting measure via its moment sequence whenever the problem is well-posed. Typically, such eigenvalue density functions are compactly supported, in which case the moments uniquely determine the distribution. For single-trace models, these moments can be computed using Schwinger–Dyson equations. For multitrace models, the moment equations couple nonlinearly, and the bootstrap
approach introduced in later sections provides one of the few systematic methods for obtaining
rigorous bounds.

The study of Dirac ensembles therefore poses two distinct mathematical challenges:
\begin{enumerate}
\item the absence of standard integrability techniques for multitrace and multimatrix models, and
\item the need to determine large-\(N\) spectral data without closed-form solutions.
\end{enumerate}
Both challenges motivate the use of positivity constraints and bootstrap methods as a substitute for
exact solvability.

% ================================
\section{Bootstrapping Dirac Ensembles via Positivity}
\label{sec:bootstrap-dirac}

Bootstrapping has re-emerged in recent years in a mathematical context quite far from its original domain in particle physics. The first appearance in a random matrix setting was in the work of Henry Lin \cite{bootstraps}, who formulated positivity-constrained spectral conditions for matrix ensembles. Further developments have been carried out in joint work of the authors and collaborators \cite{Khalkhali2022,bootstraps critical behaviour}, the  work of Kazakov and Zheng \cite{bootstraps Kazakov}, and the more recent work of Li and collaborators \cite{bootstraps Li analytic trajectory,bootstraps Li Yang-Mills matrix integrals}. Besides the work on matrix integrals, bootstrapping with positivity has been used in matrix quantum mechanics \cite{bootstraps berenstein,bootstraps matrix quantum mechancis cho}, lattice Yang-Mills theory \cite{bootstraps Guo lattice yang-mills}, lattice gauge theory\cite{Anderson,bootstraps Li lattice gauge}, and many more areas of theoretical physics. The extent of its widespread use warrants a review of its own.  

\medskip
Details of these constructions will be presented in this Section. The key point is that, just as in the conformal bootstrap, one imposes positivity and symmetry constraints on correlation data, now encoded in matrix moments or eigenvalue distributions, and derives sharp bounds on allowed spectral quantities.

The analytic study of multitrace and multimatrix Dirac ensembles is obstructed by the lack of closed-form
methods such as orthogonal polynomials or topological recursion. The bootstrap approach provides an
alternative route: instead of solving the model exactly, one derives consistency equations for the correlators
and imposes positivity constraints that must be satisfied by any admissible solution. The method consists of
four steps:

\begin{enumerate}
\item Derive the Schwinger--Dyson equations (SDE's) and rewrite them in terms of polynomial relations among
      moments, often called \emph{loop equations} in the large $N$ limit.
\item Determine the dimension of the search space, i.e., the linear span of independent moments needed
      to generate all higher moments. Note that this space need not be finite-dimensional.
\item Construct the Hankel (or moment) matrix indexed by words in the matrix variables, and impose that
      it is positive semidefinite.
\item Choose a cutoff $\Lambda$ and restrict attention to the tracial moments of words of length $\leq \Lambda$ in the alphabet of matrix variables; the resulting
      semidefinite feasibility problem yields bounds on the unknown moments.
\end{enumerate}

This setup parallels the numerical conformal bootstrap, where crossing symmetry provides the linear relations
and unitarity provides the positivity constraints; here the SDE's play the role of the crossing equations and
the moment matrix replaces the conformal block matrix.

In practice, once one determines all moments, the density function of the measure can be recovered by taking the appropriate integral transform of the characteristic function or some generating function of all moments. However, in practice the bootstrap method is only used to approximate a finite set of moments. Such approximations could be used to conjecture the general formulae for moments, such as in an example we will discuss in a later section. Alternatively, it is possible to reconstruct an approximation to the density function given a sufficiently large enough sample of moments via numerical methods. The recent work of Kováčik, and Magdolenová reconstructs the eigenvalue distribution and free energy of several matrix models from bootstrapped data \cite{bootstraps reconstruction}.

\subsection{The Hamburger Moment Problem and Hankel Positivity}

The mathematical foundation of the bootstrap method in random matrix theory is the
classical \emph{Hamburger moment problem}: given a sequence of real numbers
\[
(m_0, m_1, m_2, \dots),
\]
does there exist a positive Borel measure $\mu$ on $\mathbb{R}$ such that
\[
m_k = \int_{\mathbb{R}} x^k \, d\mu(x), \qquad k = 0,1,2,\dots \, ?
\]
The key result is that \emph{moment realizability is equivalent to the
positive semidefiniteness of an infinite Hankel matrix}.  Define the Hankel matrix
\[
\mathcal{H} =
\begin{pmatrix}
m_0 & m_1 & m_2 & m_3 & \cdots \\
m_1 & m_2 & m_3 & m_4 & \cdots \\
m_2 & m_3 & m_4 & m_5 & \cdots \\
m_3 & m_4 & m_5 & m_6 & \cdots \\
\vdots & \vdots & \vdots & \vdots & \ddots
\end{pmatrix}.
\]
The fundamental theorem (Hamburger, 1920, cf. \cite{AKHZ})   states that 
\[
\exists\,\mu \ge 0 \;\; \Longleftrightarrow \;\; \mathcal{H} \succeq 0,
\]
meaning that all principal minors of $\mathcal{H}$ are nonnegative.  The implication
``$\Leftarrow$'' follows from the fact that for every polynomial
$p(x) = \sum_{k=0}^n a_k x^k$,
\[
\sum_{i,j=0}^n a_i\, m_{i+j}\, a_j
\;=\;
\int_{\mathbb{R}} |p(x)|^2\, d\mu(x) \;\ge\; 0,
\]
so $\mathcal{H}$ acts as a Gram matrix with respect to the $L^2(\mu)$ inner product.  By a theorem of Carleman (\cite{AKHZ}) if the moments satisfy
\[
\sum_{k=1}^{\infty} m_{2k}^{-\frac{1}{2k}} = \infty,
\]
then the representing measure $\mu$ is unique.

\medskip

\paragraph{Example: the Gaussian distribution.}
Let $\mu$ be the standard normal measure
\[
d\mu(x) = \frac{1}{\sqrt{2\pi}} e^{-x^2/2}\,dx.
\]
Its moments are
\[
m_{2k} = (2k-1)!! = 1 \cdot 3 \cdot 5 \cdots (2k-1), \qquad m_{2k+1} = 0.
\]
The truncated Hankel matrix of order $4$ is
\[
\mathcal{H}_4 =
\begin{pmatrix}
1 & 0 & 1 & 0 \\
0 & 1 & 0 & 3 \\
1 & 0 & 3 & 0 \\
0 & 3 & 0 & 15
\end{pmatrix}.
\]
Its leading principal minors are
\[
\det(1) = 1,\qquad
\det\!\begin{pmatrix}1&0\\0&1\end{pmatrix} = 1,\qquad
\det\!\begin{pmatrix}1&0&1\\0&1&0\\1&0&3\end{pmatrix} = 2,\qquad
\det(\mathcal{H}_4)= 12,
\]
all strictly positive, confirming realizability.  In fact, every finite truncation
$\mathcal{H}_n$ of the Gaussian moment matrix is positive definite, reflecting the
determinate (unique) nature of the Gaussian moment problem. Any  Gaussian/normal distribution is in fact uniquely defined by its first two moments or equivalently first two cumulants. This fact can be readily observed from the form of its characteristic function.

\medskip

In random matrix models, one studies the limiting spectral measure
\[
m_k = \lim_{N\to\infty} \frac{1}{N}\, \mathbb{E}[\mathrm{Tr}(H^k)],
\]
and the requirement that this define a legitimate probability measure on $\mathbb{R}$
imposes the \emph{Hankel positivity constraint}
\[
\mathcal{H}(m_0,m_1,\dots,m_{2n}) \succeq 0.
\]
In bootstrap applications, the moment sequence is not known a priori; instead,
one solves a system of algebraic relations (Schwinger--Dyson equations) subject
to the Hankel positivity condition.  The feasible region in moment space is thus
cut out by positivity of a semidefinite Hankel matrix, in direct analogy with the
unitarity constraints of the conformal bootstrap.

In the context of Dirac ensembles, the situation is similar but the moments arise from tracial moments of the Hermitian matrices and 
of the Dirac operator itself:
\[
d_\ell
  \;=\;
  \lim_{N\to\infty}
  \frac{1}{N^2 Z}
  \int_{\mathcal{D}_{(p,q)}} \mathrm{Tr}(D^\ell)\, e^{-S(D)}\, dD,
\]
where $\mathcal{D}_{(p,q)}$ denotes space of admissible Dirac operators associated to fuzzy geometries of a fixed signature. Just as before, these numbers assemble into a Hankel matrix
\[
\mathcal{D}
=
\begin{bmatrix}
1   &  d_1 &  d_2 & d_3 & \cdots\\
d_1 &  d_2 &  d_3 & d_4 & \cdots\\
d_2 &  d_3 &  d_4 & d_5 & \cdots\\
d_3 &  d_4 &  d_5 & d_6 & \cdots\\
\vdots & \vdots & \vdots & \vdots & \ddots
\end{bmatrix}.
\]
Since $\mathrm{Tr}(P^\ast P)\ge0$ for all noncommutative polynomials $P(D)$, this matrix must be positive
semidefinite for any admissible Dirac ensemble. The bootstrap procedure therefore imposes
\[
\mathcal{D} \;\succeq\; 0
\]
together with the loop equations of Section~\ref{sec:bootstrap-dirac}. The intersection of these two
constraint sets defines the allowed region for the unknown moments.

In practice, one truncates the Hankel matrix to moments of degree $\le \Lambda$, yielding a finite-dimensional
semidefinite program. The feasible set in the $(d_1,d_2,\dots)$-space then gives rigorous upper and lower
bounds on the low-degree moments as functions of the coupling constant $g$ in the action. For example, using
only moments of order $\le4$ already produces nontrivial bounds on the second moment $m_2$ of the associated
Hermitian matrix model.

% ================================
\subsection{Schwinger--Dyson  Equations}

Let $\mathcal{H}_N$ denote the space of $N\times N$ Hermitian matrices, and consider a general multimatrix,
multitrace model with $m$ matrix variables:
\[
Z \;=\; \int_{\mathcal{H}_N^m} \exp\!\bigl( - S(H_1,\dots,H_m) \bigr)\, dH_1\cdots dH_m,
\]
where $S$ is a polynomial in the matrices and their traces. For any word
$W = H_{i_1} H_{i_2} \cdots H_{i_\ell}$ in the alphabet $\{H_1,\dots,H_m\}$, the corresponding moment is defined
by
\[
\ex \left[ \frac{1}{N}\mathrm{Tr}(W) \right]
  \;=\;
  \frac{1}{N Z}
  \int_{\mathcal{H}_N^m}
     \mathrm{Tr}(W)\, e^{-S(H_1,\dots,H_m)}\, dH_1\cdots dH_m.
\]
The Schwinger--Dyson equations arise from the identity
\[
0 \;=\; \sum_{a,b=0}^{N}\int_{\mathcal{H}_{N}^{m}} \frac{\partial}{\partial H_{a b}^{(i)}}\Bigl( W_{b a}\, e^{-S(H_{1},...,H_{m})} \Bigr)dH_{1}\cdots dH_{m},
\]
and yield relations among correlators. In the simplest case of a one-matrix model with $W=H^\ell$ and $S(H)= \frac{N}{2}\tr H^{2}- \sum_{j\geq 3}^{d} \frac{N t_{j}}{j}^{d} \tr H^{j}$,
one obtains
\[
\sum_{k=0}^{\ell-1}
  \ex \left[ \mathrm{Tr}\bigl( H^{\ell-1-k} \bigr)\,
              \mathrm{Tr}\bigl( H^{k} \bigr) \right]
  \;=\;
  N \ex\left[ \mathrm{Tr}\bigl( H^{\ell}\, S'(H) \bigr) \right].
\] 

By taking the large $N$ limit, the leading-order factorization property
\[
\ex \left[ \frac{1}{N}\mathrm{Tr}(H^a)\, \frac{1}{N}\mathrm{Tr}(H^b) \right]
  \;=\;
  \ex \left[ \frac{1}{N}\mathrm{Tr}(H^a) \right]
  \ex \left[\frac{1}{N}\mathrm{Tr}(H^b) \right]
  \;+\; O(N^{-2})
\]
allows the SDE's to be written solely in terms of the limiting single tracial moments:
\[
m_k \;=\; \lim_{N\to\infty} \frac{1}{N}\ex \left[ \mathrm{Tr}(H^k) \right],\qquad k=1,2,\dots\,.
\]

When the model is treated as a formal matrix integral the factorization of moments follows from the genus expansion of moments, see for example Appendix of \cite{Khalkhali2022}. Otherwise, for convergent integrals, the factorization of moments can be proven using properties of the convergence of the empirical eigenvalue distribution \cite{Johansson,Khalkhali2025 large N limit}. In the above single  single trace one-matrix model the loop equations for $\ell\geq 0$, can be written as 
\begin{equation*}
	m_{\ell+1} = \sum_{k=0}^{\ell-1}m_{\ell-1-k}m_{k} +\sum_{j\geq 3} t_{j} m_{\ell+j-1}. 
\end{equation*}
As we will see in subsequent sections, for multi-tracial models additional products of moments will appear.

% ================================
\subsection{Search Space and Positivity Constraints}

Given the loop equations, the next task is to determine the \emph{search space}: the smallest finite-dimensional
vector space of low-degree moments that linearly generates all other moments through the SDE's. For instance, in the example of the one-matrix single trace model, all of the loop equations, can be written in terms of $m_{1},m_{2},...,m_{d-2}$, and the coupling constants, so the dimension of the search space is said to be $d-2$. In multi-matrix models it is in general much more difficult to determine the search space dimension.

%For the Dirac ensembles arising from finite spectral triples, a notable simplification occurs: although the actions are multitrace and multimatrix, the resulting moment space is one-dimensional. This mirrors the situation in classical one-matrix models, but is far from obvious a priori.

Once the search space is known, one forms the Hankel (moment) matrix
\[
M_{\Lambda} = \bigl( m_{u^\ast v} \bigr)_{u,v},
\]
indexed by all words $u,v$ in the matrix variables of length $\leq \Lambda$. By construction, this matrix
must be positive semidefinite, since for any noncommutative polynomial $P$ one has
\[
\ex \left[\frac{1}{N}\mathrm{Tr}\bigl( P^\ast P \bigr) \right] \;\geq\; 0.
\]
Together, the loop equations and the positivity of $M_\Lambda$ define a semidefinite feasibility problem,
whose solution space consists of all possible large-$N$ moment sequences consistent with the model.
In this way, one obtains rigorous upper and lower bounds on the moments without solving the model exactly.

The bootstrap method therefore provides a non-perturbative analytic tool for Dirac ensembles, playing the
same role that conformal bootstrap methods play in CFT: instead of solving the theory, one cornerstones
the space of possible solutions using symmetry and positivity alone.

So far all works that the authors are aware of have restricted their attention to bootstrapping Hermitian matrix integrals in the large $N$ limit. The factorization of moments discussed in the previous section allows for the above Hankel matrix to be the only required constraints to solve the system. If one were to consider bootstrapping matrix ensembles for finite $N$ or even subleading contributions, additional constraints woudl be required to be derived from Hankel matrices with multi-tracial moments.

\subsection{Example: The Cubic Type $(1,0)$ Dirac Ensemble}

A simple, yet analytically rich model arises when the action is
cubic. In the signature $(1,0)$ case, the Dirac operator
reduces to a single Hermitian matrix variable $H$, and the partition function
takes the form
\begin{align*}
	Z &= \int_{\mathcal{D}} \exp\!\left(-\frac{1}{4}\,\mathrm{Tr}(D^{2})
	- \frac{g}{6}\,\mathrm{Tr}(D^{3})\right) \, dD \\
	&= \int_{\mathcal{H}_{N}} \exp\left(-\frac{1}{2}\left(N \operatorname{Tr} H^2+(\operatorname{Tr} H)^2\right)-\frac{g}{3}\left(N \operatorname{Tr} H^3+3 \operatorname{Tr} H^2 \operatorname{Tr} H\right)\right)dH.
\end{align*}

\begin{figure}[H]
	\centering
		\includegraphics[width=.5\textwidth]{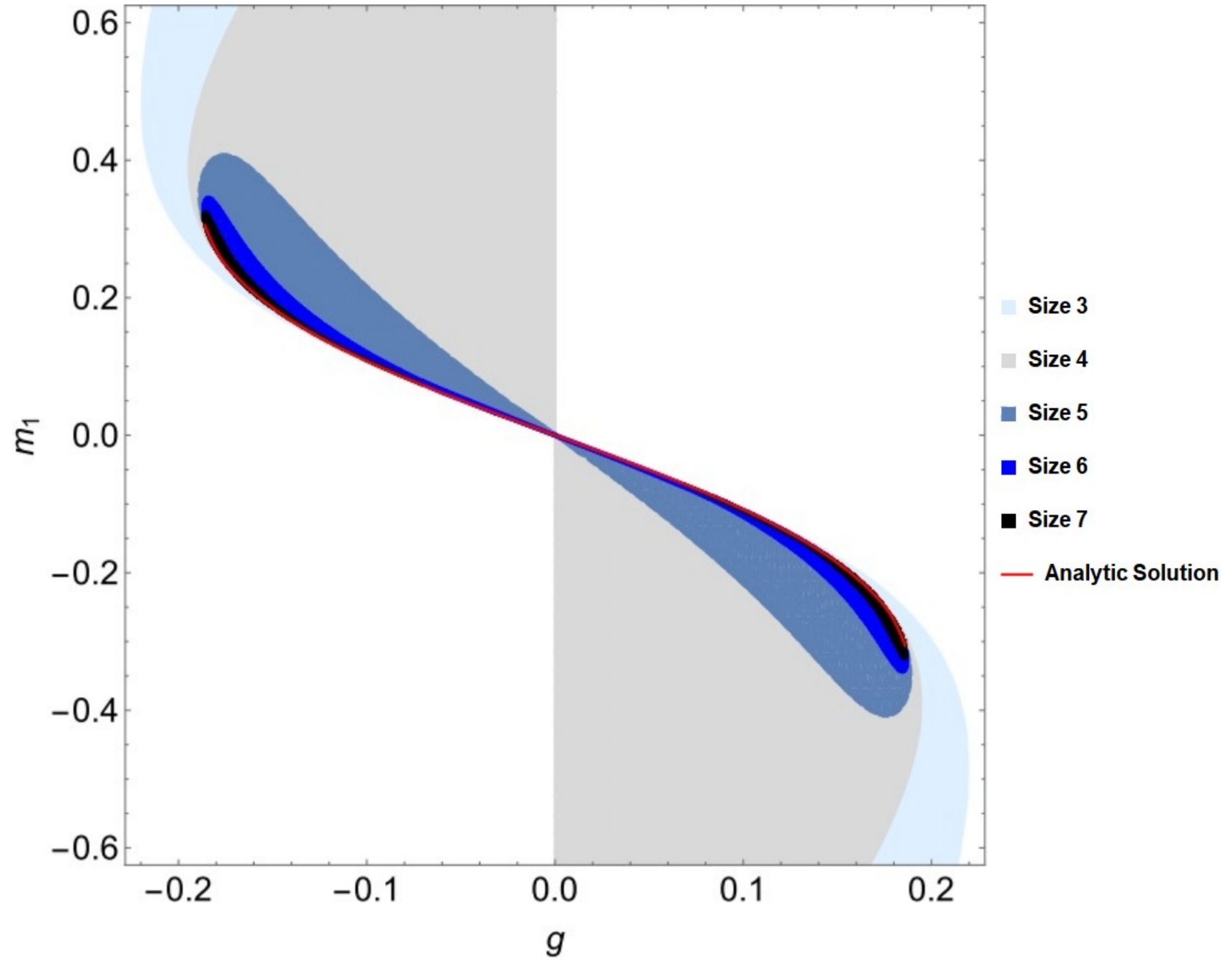}
	\caption{Bootstrapped regions of possible solutions to the loop equations for the cubic type $(1,0)$ Dirac ensemble\cite{hessam2022fromnoncom}. Each colour corresponds to a different number of constraints derived from positivity of some principal
		minors of the specified size. The analytic solution found in \cite{hessam2023double} is plotted in red for comparison.}
	\label{fig:cubic}
\end{figure}

This model is not convergent for any nonzero value of $g$, so must be interpreted as a formal matrix integral. For $\ell\geq 0$, the loop equations of the model are of the form

$$
\sum_{k=0}^{\ell-1} m_k m_{\ell-k-1}=m_{\ell+1}+m_1 m_{\ell}+g\left(m_{\ell+2}+2 m_1 m_{\ell+1}+m_2 m_{\ell}\right).
$$
From the loop equations it is clear that all moments can be written in terms of $m_{1}$, so the search space has dimension one. The bootstrap method provides a way to determine the allowed
pairs $(g,m_{1})$ in the large $N$ limit by imposing positivity of the
(noncommutative) moment matrix associated with the tracial sequence
$(m_{0},m_{1},m_{2},\dots)$. By enforcing positivity of progressively larger
principal minors, one obtains a nested family of feasible regions in the
$(g,m_{1})$--plane. As the number of imposed constraints increases, these
regions shrink, revealing a nonlinear relation between the coupling $g$ and
the first moment $m_{1}$. See Figure \ref{fig:cubic}.

Where the analytic solution stops near $\pm 0.2$, are two critical points where the free energy fails to be analytic in $g$. The critical phenomenon around these points has been related to the $(3,2)$ minimal model from Liouville quantum gravity \cite{hessam2023double}. In general, bootstrapping with positivity can be used to estimate the location of a critical point as well as its critical exponents. See for example the recent work of the authors and collaborators \cite{bootstraps critical behaviour}.

% \begin{
\subsection{Example: the Quartic Type $(1,0)$ and $(0,1)$ Dirac Ensembles}	
  The type $(1,0)$ and $(0,1)$ fuzzy geometries have Dirac operators of the form 
$$D_{(1,0)} = \{H,\cdot\}  \quad \text{and}\quad D_{(0,1)} = [H,\cdot],$$
respectively, where $H$ is Hermitian $N$ by $N$. The Dirac quartic Dirac ensemble considered is
\begin{equation*}
	\frac{1}{Z} e^{-g\tr D^2 -\tr D^4}dD.
\end{equation*}
In terms of $H$, it can be written as 
\begin{equation} \label{action}
	\frac{1}{Z} \exp\{-2g(N\tr H^2 +\varepsilon(\tr H)^2) -2(N \tr H^{4} +\varepsilon 4 \tr H \tr H^3 + 3(\tr H^2)^2\}dH,
\end{equation}
where $\varepsilon =1$ for type $(1,0)$ and $\varepsilon =-1$ for type $(0,1)$. In the case of the type $(0,1)$ geometry there are some additional nuances to defining the integral, see \cite{Khalkhali2025 large N limit} for details. 

The loop equations are of the form:
\begin{equation*}
	\sum_{k=0}^{\ell-1} m_k m_{\ell-k-1}=t_{2}\left(4 m_{\ell+1}+ 4 \epsilon  m_1 m_{\ell}\right)+8m_{\ell+3}+ 8\epsilon  m_3 m_{\ell}+ 24\epsilon m_1 m_{\ell+2}+16 m_2 m_{\ell+1}.
\end{equation*}
Note that by the symmetric of the matrix integral $H \rightarrow -H$, all odd moments are zero when considering either a formal or convergent matrix integral. As a consequence, both models have the same loop equations since all terms with an $\epsilon$ are zero. From the above formula it is clear that all moments can be written in terms of $m_{2}$, so the search space dimension is one.

\begin{figure}[H]
	\centering     %%% not \center
	\subfigure[]{\label{fig:a}\includegraphics[width=60mm]{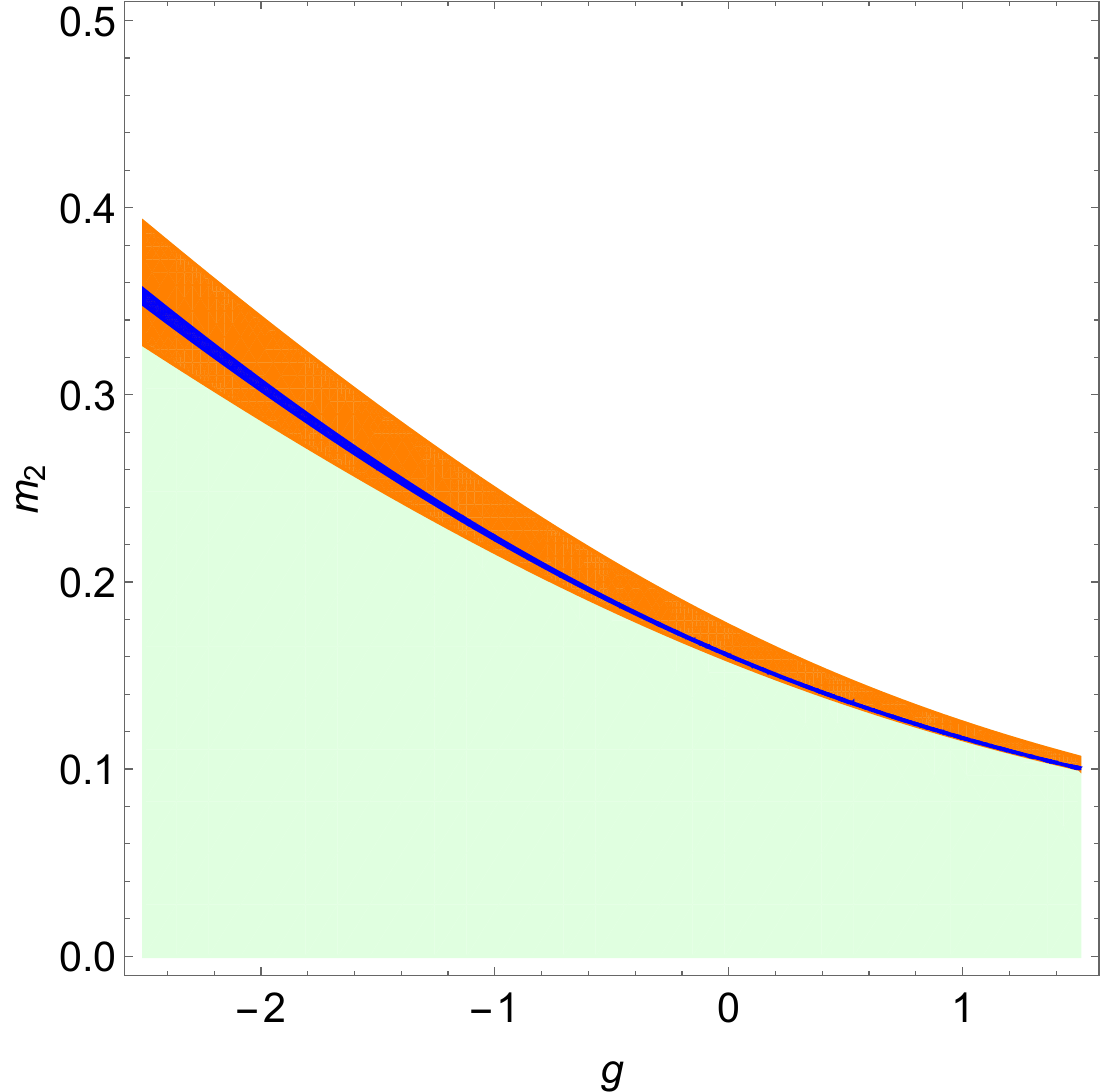}}
	\subfigure[]{\label{fig:b}\includegraphics[width=60mm]{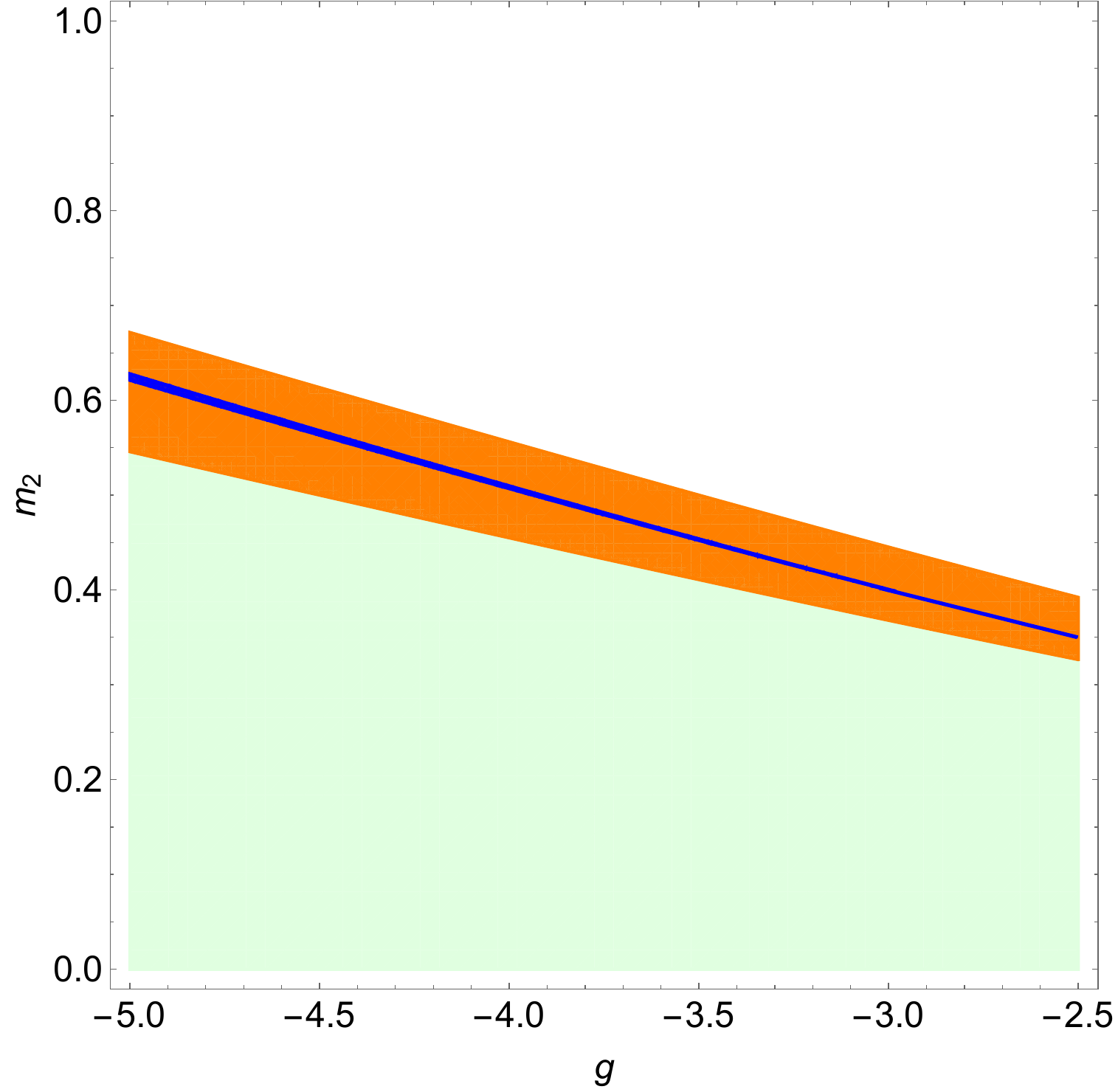}}
	\caption{Bootstrapped region of solution space for the type $(1,0)$ quartic Dirac ensemble \cite{hessam2022bootstrapping}. The different
		coloured regions denote different combinations of constraints applied.}
	\label{fig:asymmetric (0,1)}
\end{figure}
The analytic solution was found in  \cite{Khalkhali2020} for the convergent case and \cite{hessam2023double} for the formal case. These models also have critical points like the cubic model, however, since we only have a coupling in front of the Gaussian term, it is not immediately visible until one rescales $D$ and $g$ to include a redundant coupling in front of $\tr D^4$ in the potential.  

\subsection{Asymmetric Solutions}
By the symmetry of the quartic type $(1,0)$ and $(0,1)$ ensmebles, for either formal or convergent matrix integral, all odd moments must be equal to zero. However, what is interesting is that if one does not restrict all odd moments to being zero, asymmetric solutions appear when bootstrapping. What is even more remarkable is that bootstrapped solution of type $(0,1)$ case still only produces symmetric solutions, while the type $(1,0)$ produces genuine asymmetric solutions. All moments can be written in terms of $m_{1}$ and $m_{2}$, so the search space has dimension two. See Figure \ref{fig:asymmetric (1,0)} and Figure \ref{fig:asymmetric (0,1)} for the results. These phenomena will be investigated in an upcoming work \cite{Bukor2026} and serves as an interesting example of a bootstrapping problem with multiple solutions. The authors believe that more than two asymmetric solutions are possible, so if given sufficient computational resources we expect the asymmetric subspaces of the solution space to fork again.  

\begin{figure}[H]
	\centering
	\includegraphics[width=.5\textwidth]{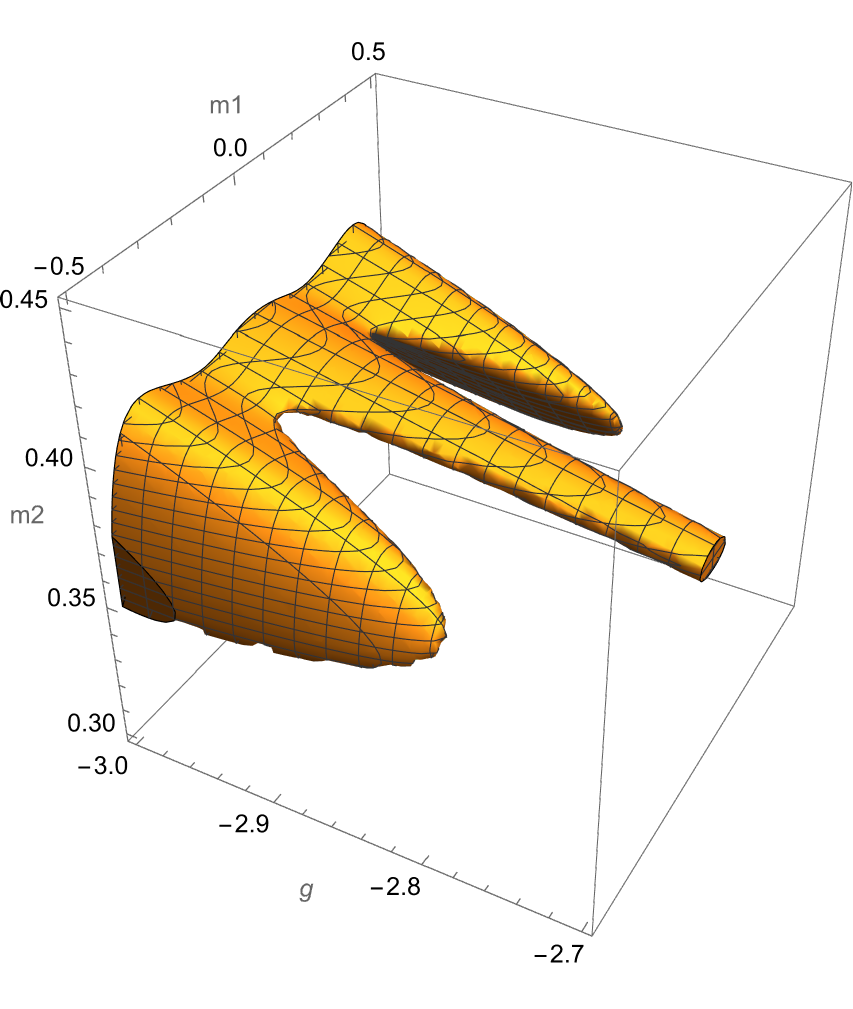}
	\caption{Bootstrapped region of solution space for the type $(1,0)$ quartic Dirac ensemble, allowing for asymmetric solutions. This plot was generated using ten SDE's and a submatrix of the Hankel matrix of size ten. There is a clear fork in the solutions space, where each outer prong corresponds to a subspace of possible asymmetric solutions. }
	\label{fig:asymmetric (1,0)}
\end{figure}

\begin{figure}[H]
	\centering     %%% not \center
	\subfigure[]{\label{fig:a2}\includegraphics[width=60mm]{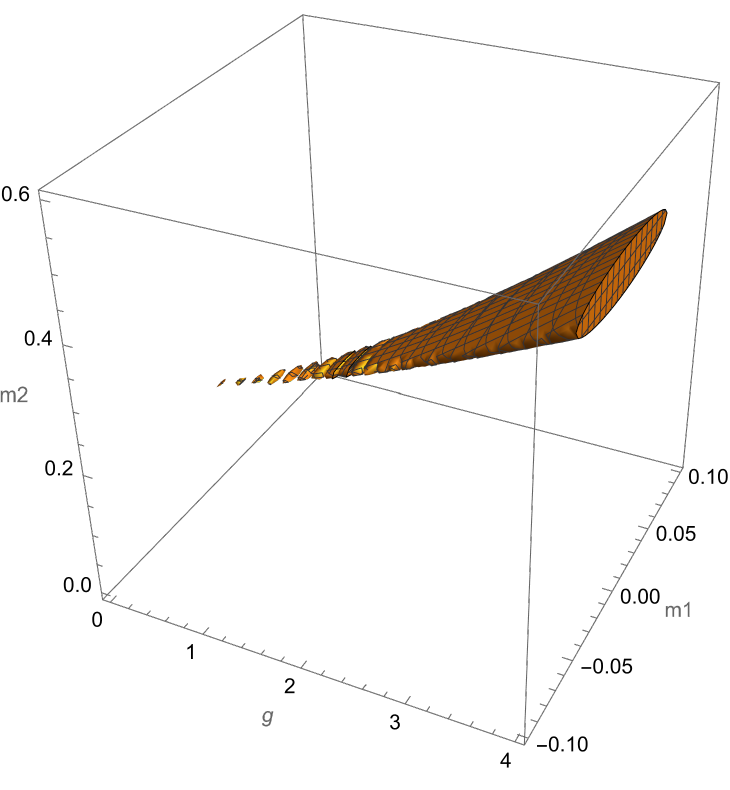}}
	\subfigure[]{\label{fig:b2}\includegraphics[width=60mm]{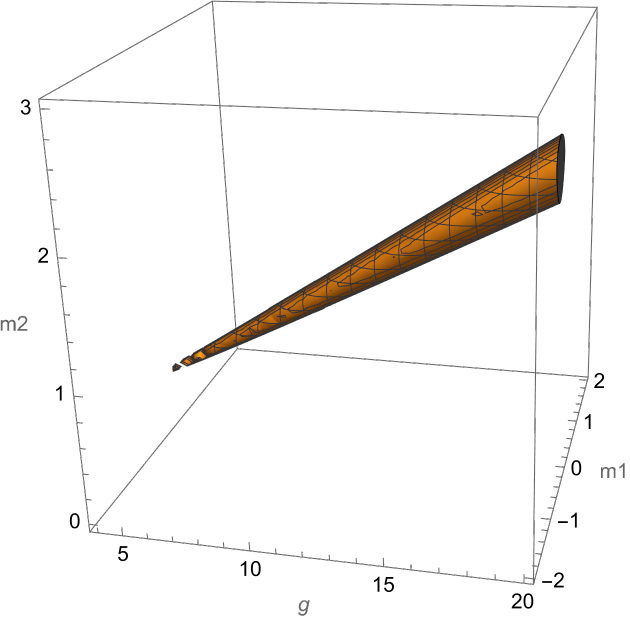}}
	\caption{Bootstrapped region of solution space for the type $(0,1)$ quartic Dirac ensemble, allowing for asymmetric solutions. These were found using eight SDE's and a submatrix of the Hankel matrix of size 8.}
		\label{fig:asymmetric (0,1)}
\end{figure}

Such solutions as mentioned before cannot belong to any matrix model as discussed in this paper. It is an open question to find what type of mathematical object asymmetric solutions to the SDE's represent.  It is also worth noting that asymmetric solutions of even models do not existence in single trace matrix models since and are in fact a result of multi-tracial terms. Such asymmetric solutions appear in fuzzy field theories \cite{Lizzi,Samuel}, and were studied in  the recent work \cite{Bukor asymmetric}.

\subsection{Type (2,0) Dirac Ensembles}

The Dirac operator
of type $(2,0)$ Dirac ensemble can be written in block
form in terms of two Hermitian $N \times N$ matrices $A$ and $B$:
\begin{equation*}
	D =\left[\begin{array}{cc}
		1 & 0 \\
		0 & -1
	\end{array}\right] \otimes\{A, \cdot\}+\left[\begin{array}{ll}
	0 & 1 \\
	1 & 0
\end{array}\right] \otimes\{B, \cdot\} .
\end{equation*}
The 
spectral action then becomes a multitrace, multimatrix
polynomial in $A$ and $B$. For instance, one computes
\[
\mathrm{Tr}(D^{2}) = 4N \, \mathrm{Tr}(A^{2}) + 4N \, \mathrm{Tr}(B^{2})
   + 4(\mathrm{Tr}\,A)^{2} + 4(\mathrm{Tr}\,B)^{2},
\]
and the quartic term expands as
\begin{align*}
\mathrm{Tr}(D^{4}) &= 4N \, \mathrm{Tr}(A^{4}) + 4N \, \mathrm{Tr}(B^{4})
 + 16N \,\mathrm{Tr}(A^{2}B^{2}) - 8N \,\mathrm{Tr}(ABAB)  \\
&\quad + 16\,\mathrm{Tr}(A)\,\mathrm{Tr}(A^{3})
     + 16\,\mathrm{Tr}(A)\,\mathrm{Tr}(B^{2}A)
     + 16\,\mathrm{Tr}(B)\,\mathrm{Tr}(B^{3})  \\
&\quad + 16\,\mathrm{Tr}(B)\,\mathrm{Tr}(A^{2}B)
     + 16\,(\mathrm{Tr}\,AB)^{2}
     + 12\,(\mathrm{Tr}\,A^{2})^{2}
     + 12\,(\mathrm{Tr}\,B^{2})^{2} \\
&\quad + 8\,\mathrm{Tr}(A^{2})\,\mathrm{Tr}(B^{2}).
\end{align*}
Let us take the partition function for the $(2,0)$ Dirac ensemble as
\[
Z = \int_{\mathcal{D}_{(p,q)}} e^{-t_{2}\tr D^2 - t_{4} \tr D^4}dD. 
\]
The resulting model is therefore a multimatrix multitrace random
matrix ensemble, lying well outside the usual one-matrix polynomial potential
framework of classical random matrix theory. In \cite{Khalkhali2024 coloured maps},  a solution in the large $N$ limit of the quartic type $(2,0)$, $(1,1)$ and $(0,2)$ Dirac ensembles was proposed and proved using the bootstrapped solution found in \cite{hessam2022bootstrapping} as guidance. Unfortunately, later an error was found in the proof. However, the result is still conjectured to be true based off of bootstrapped data. It is still true and was proven in \cite{Khalkhali2025 large N limit} that any unique symmetric solution is the same for all three signatures since the loop equations are the same. For simplicity we have only discussed the the type $(2,0)$ case.

{\bf Conjecture:}
	The second moment of the formal and convergent type $(2,0)$ quartic Dirac for $t_{2}$ and $t_{4}$ in a sufficiently small enough neighbourhood of zero, can be written as 
	$$\lim_{N \rightarrow \infty} \frac{1}{N} \mathbb{E}[\tr D^2]= 
	\frac{1}{8 t_{4}} \left({\sqrt{{t_2}^2+8
			{t_4}}}-{{t_2}}\right).$$

It was proven that if a solution exists that entire hierarchy of moments can be expressed recursively in terms of these
quantities, showing that the algebra of relations among moments is
one-dimensional in the large $N$ limit assuming there is indeed a solution. Once the potential is rescaled, the conjectured solution has a critical point whose critical exponent corresponds to the university class of the Continuum Random Tree.

To study the large $N$ limit, we introduce the notation
\[
m_{a,b,c,d} \;=\; \lim_{N\to\infty}
   \ex \left[ \mathrm{Tr}\bigl(A^{a} B^{b} A^{c} B^{d}\bigr) \right],
\]
and similarly write $m_{k}$ for single--matrix moments such as
$\mathrm{Tr}(A^{k})$ or $\mathrm{Tr}(B^{k})$ where no ambiguity exists since the action is symmetric in $A$ and $B$.
Applying the Schwinger--Dyson equations to words of the form
$A^{\ell}$ and using cyclicity of the trace, one obtains the SDE's
\begin{align*}
\sum_{k=0}^{\ell-1} m_{k}\, m_{\ell-k-1}
  &= (8g + 64 m_{2})\, m_{\ell+1} + 16\, m_{\ell+3} \\
  &\quad - 16\, m_{\ell,1,1,1}
     + 32\, m_{\ell+1,2},
\end{align*}

For more examples of the loop equations and moments formulae based on the conjecture, see the Appendices of \cite{Khalkhali2024 coloured maps}.

\begin{figure}[H]
	\centering     %%% not \center
	\subfigure[]{\label{fig:a1}\includegraphics[width=60mm]{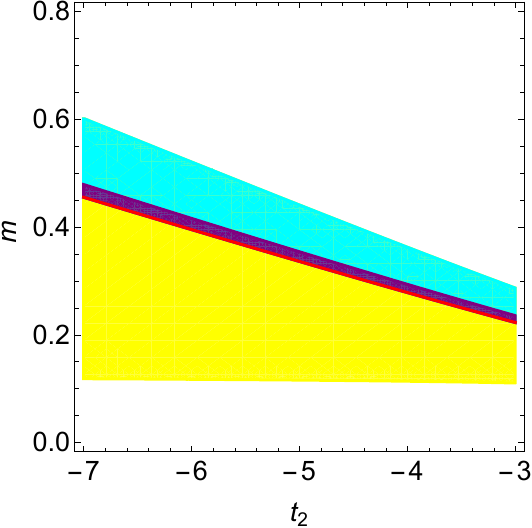}}
	\subfigure[]{\label{fig:b1}\includegraphics[width=60mm]{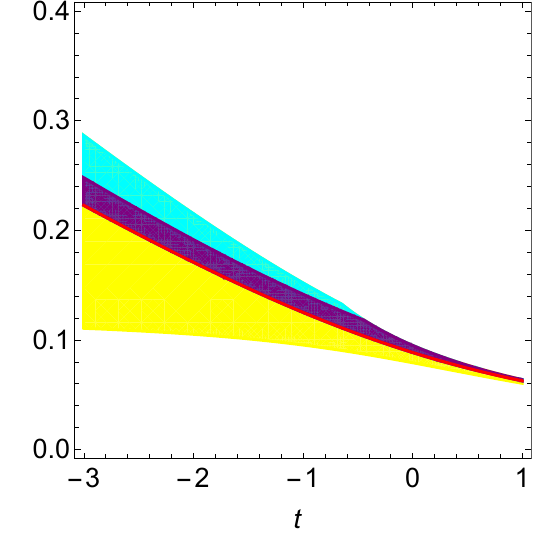}}
	\caption{Bootstrapped region of solution space for the type $(2,0)$ quartic Dirac ensemble \cite{hessam2022bootstrapping}. The different
		coloured regions denote different combinations of constraints applied. The red curve is the conjectured analytic solution.}
	\label{fig:(2,0))}
\end{figure}

\subsection{Random Metric Spaces and Quantum Gravity}
Single trace formal matrix models enumerate isomorphism classes graphs embedded onto surfaces of various genus, called \textit{maps}. One can construct discrete probability distributions using these generating functions of maps, and it can be shown that the random metric spaces constructed from such considerations converge in the Gromov-Hausdorff sense to random fractal surface. For a very general class of such maps the limiting surface is the Brownian map \cite{LeGall,Miermont}. The Brownian map is the prime example of a Liouville quantum gravity surface \cite{gwynne}.  For an accessible review of the construction of such objects see \cite{LeGall review} and for a review of applications to two dimensional quantum gravity and Liouville quantum gravity see \cite{Budd QG}. 

In \cite{hessam2023double}, it was shown that that single trace matrix models are non-trivially inside several type $(1,0)$ Dirac ensembles, which then implies that the associated random metric spaces converge to the Brownian map at certain critical points. What is more interesting is that there are critical points in such Dirac ensembles that have no known random metric space when tuned to criticality. Multi-tracial matrix ensembles enumerate \textit{stuffed maps}, these are a generalization of the maps such that generalize the type of embedding of the graph. They contain bridges between graph connected components which are viewed as ``wormholes between baby universes" in the physics literature and are believed ot be associated duality to Liouville quantum gravity based on heuristics \cite{Rhodes review}. In the case of the Brownian map and analogous results, the proof relies on a bijection between maps and labeled plane trees called mobiles \cite{Bouttier}. Thus, a first step towards in finding a the limiting random metric space for stuffed maps is to construct an analogous bijection which has been done in the recent work of the second author \cite{Pagliaroli2025}. Future, work will aim to establish such convergence results for stuffed maps and thus give a random metric space associated to Dirac ensembles. Note that the critical exponents of the conjectured solution to the type $(2,0)$ quartic Dirac ensemble are associated with the Random Continuum Tree which is more common in tensor models than matrix models. Based on Monte Carlo simulations is also very likely that there are other solutions to the type $(2,0)$ quartic Dirac ensemble that have different critical exponents.

\section{Bootstrapping  Multi-matrix Quantum Mechanics}
In this  section we briefly review  the recent work of Lin and Zheng
\cite{LinZheng2025}, who apply the bootstrap philosophy to multi--matrix
quantum mechanics arising from the dimensional reduction of Yang--Mills
theories.  Their goal is to determine ground state correlation functions
of large $N$ bosonic matrix models with high numerical precision by
imposing only consistency, symmetry, and positivity.

The authors consider  traceless Hermitian matrices (with noncommutative operator entries)
\[
X^I, \qquad P^I, \qquad I=1,\dots, D,
\]
transforming in the adjoint of $SU(N)$, such that 
\begin{equation*}
	[(X_{I})_{ij},(P_{J})_{k\ell}] = \sqrt{-1} \delta_{i \ell}\delta_{j k} \delta_{I J}.
\end{equation*}
The Hamiltonian is defined as
\begin{equation}
H \;=\; 
\frac{1}{2} \sum_{I=1}^D \operatorname{Tr}\!\big( P^I P^I \big)
\,+\, \frac{M^2}{2} \sum_{I=1}^D \operatorname{Tr}\!\big( X^I X^I \big)
\,-\, \frac{g_{\mathrm{YM}}^2}{4}
\sum_{I,J=1}^D \operatorname{Tr}\!\Big( [X^I, X^J]^2 \Big).
\label{eq:Hamiltonian-LZ}
\end{equation}
The massless case $M=0$ corresponds to the bosonic part of the BFSS matrix 
model \cite{BankFiscShenSuss}.  In the large $N$ limit correlators factorize, for example
\[
\ex \left[ \tfrac1N \operatorname{Tr}(X^I X^J)\,
       \tfrac1N \operatorname{Tr}(X^K X^L) \right]
\;\xrightarrow[N\to\infty]{}\;
\ex \left[ \tfrac1N \operatorname{Tr}(X^I X^J) \right]
\ex \left[ \tfrac1N \operatorname{Tr}(X^K X^L) \right].
\]

\medskip

\noindent

Let $W$ denote a word in the $D$ matrices $X^I$.  Define the large $N$
moment
\[
m_W := 
\lim_{N\to\infty} 
\frac{1}{N}\ex \left[ \operatorname{Tr} W(X^1,\dots,X^D)
\right].
\]
The basic observable of interest is the second moment
\[
m_2 = \frac{1}{D}\sum_{I=1}^D m_{X^I X^I},
\]
which physically corresponds to the ``size'' of the matrix under
confinement.  Higher moments such as $m_{4}$, $m_{6}$, and mixed moments
$m_{X^I X^J X^K X^L}$ also enter into the bootstrap constraints.

\medskip

\noindent

Insertions of matrix polynomials into the path integral yield
Schwinger--Dyson equations.  For example, for a word $W$ one obtains
schematically
\begin{equation}
\sum_{\substack{W = U(I)V \\ \text{splitting}}}
m_{U}\,m_{V}
\;=\;
\sum_{J} m_{\,\partial_J W}
\;-\;
g_{\mathrm{YM}}^2 
\sum_{J,K} m_{\,\partial_{J,K} W},
\label{eq:SDE-general}
\end{equation}
where the right-hand side comes from the variation of the action
(\ref{eq:Hamiltonian-LZ}) above.  In practice one truncates to all words of
length $\ell \le L$ for some bootstrap level $L$ (Lin--Zheng push this to
$L=14$).

\medskip

\noindent

The crucial bootstrap input is positivity of the ground state $\Omega$:
for any operators $O_i$,
\[
\langle \Omega \,|\, O_i^\dagger O_j \,|\, \Omega\rangle \ge 0.
\]
Choosing $O_i$ to be all words of length $\le L$ yields a positive
semidefinite moment matrix,
\begin{equation}
\mathcal{M}_{ij}
=
\langle \Omega \,|\, W_i^\dagger W_j \,|\, \Omega\rangle
\;\succeq\; 0.
\label{eq:moment-matrix}
\end{equation}
This is the analogue of the Hankel matrix in the classical Hamburger
moment problem.

\medskip

\noindent

Now the system consisting of
all SDE's (\ref{eq:SDE-general}) for words of length
$\le L$ and the
positivity of constraints from the moment matrix (\ref{eq:moment-matrix}). These 
defines an admissible region in the space of moments (notably the
parameter $m_2$).  
Lin and Zheng solve this system using nonlinear relaxation methods to
obtain precise two-sided bounds.  At each level $L$ the allowed region
shrinks, and the authors observe rapid convergence.

\medskip

\noindent
To give an example, let us consider the  Bosonic BFSS  model ($D=9$, $M=0$).
At bootstrap level $L=10$, the allowed region for $m_2$ collapses to a
narrow interval matching and improving existing Monte Carlo estimates.
Their main graphical output is a ``bootstrap island'':
it contains all consistent solutions and shrinks as the level increases.

\medskip

\noindent

This work demonstrates that large $N$ multi-matrix quantum mechanics can be
analyzed to high precision using only:
symmetry, loop equations, and positivity.  
The technique bypasses Monte Carlo extrapolations and provides a
non-perturbative tool potentially relevant for matrix-based models of
quantum gravity such as models  studied in this paper based on noncommutative geometry.

\medskip 

There is a close relation  to Dirac ensembles and finite noncommutative geometries.
Mathematically, matrix quantum mechanics lives in the free $*$--algebra 
generated by the $X^I$, and bootstrap constraints define a 
noncommutative moment problem.  
This is directly parallel to the Dirac ensemble program, where 
finite real spectral triples $(A,H,D,J,\gamma)$ generate a 
noncommutative probability distribution for the Dirac operator $D$, and 
one studies expectations of traces of $D^k$ subject to:
\begin{itemize}
\item Clifford algebra relations,
\item KO--dimensional constraints,
\item positivity of the noncommutative measure,
\item loop equations coming from the spectral action.
\end{itemize}
In particular, the Hankel matrix for Dirac moments plays the role 
analogous to the moment matrix (\ref{eq:moment-matrix}) in the 
Lin--Zheng bootstrap.  The Barrett  classification of fuzzy 
spectral triples supplies the ``kinematic'' space in which the 
bootstrap occurs, while polynomial potentials in 
$\operatorname{Tr}(D^k)$ play the role of matrix model actions.  

Thus multi--matrix quantum mechanics and Dirac ensembles can be viewed 
as two manifestations of a single bootstrap paradigm:  
\emph{positivity + loop equations determine spectra without a Lagrangian}.  
This connection provides a conceptual justification for applying 
bootstrap ideas to the Connes--Lott and Connes--Chamseddine spectral 
action, after a suitable truncation,  and it shows that random Dirac operators can be studied using 
exactly the same positivity--based machinery that now governs modern 
large $N$ matrix mechanics.

\section{A Tale of Two Cities}

The Connes--Chamseddine spectral action principle \cite{Spectral action, QFTNCG, Neutrino Mixing} and the Dirac 
ensemble quantization program \cite{Barrett2016, hessam2022fromnoncom} both arise from the idea that geometry may be
encoded in the spectrum of a Dirac operator. Moreover, there are definite benefits
in working over a noncommutative space. However, they interpret the
Dirac operator in fundamentally different ways, and the resulting theories
have distinct mathematical structures and physical interpretations. 

It is important to note that, despite its remarkable unifying power, the
spectral action remains essentially a \emph{semiclassical} theory. The bosonic
part of the action is given by a high-energy expansion of a spectral
functional, and the fermionic action is classical. A full quantum version of
the spectral action has not yet been achieved: there is no fully established
path-integral quantization of the spectral action, and it is not clear what
the appropriate configuration space of metrics or spectral data should be in such
a quantization scheme. Several other approaches to quantizing the spectral action include \cite{vanN2022one,vanN2023,gakkhar2024spectral,Sanchez quivers,Sanchez loop equations}. Developing a genuinely quantum theory based on the
spectral action remains an open and a very interesting  problem \cite{van2024}. Related to this issue
is the fact that, in the spectral action approach to the Standard Model, 
gravity enters only through the classical Einstein--Hilbert term and its
higher-derivative corrections. Quantum gravitational degrees of freedom are, 
in a sense, absent from the framework.

The situation is almost reversed in the Dirac ensemble approach. There, the
Dirac operator itself is promoted to a dynamical random variable, and the ensemble
defines a fully quantum theory of fluctuating geometries. In this setting,
gravity is naturally quantized through the statistical distribution of spectra.
However, so far the theory we have discussed contains no fermionic degrees of freedom and there is no analogue yet of the finite noncommutative space underlying the
Standard Model. Moreover, Dirac ensembles typically describe finite
noncommutative geometries, so developing a realistic continuum limit,  i.e. one
that incorporates both fluctuating metrics and a finite internal geometry to account for fermions is
a major challenge. We will discuss efforts in recent works \cite{Barrett2024,Sanchez Yang-Mills,Khalkhali2025 large N limit} to add such features to Dirac ensembles.

In this last section we would like to compare these two applications of NCG to 
physical problems and explore whether there is any potential for merging the 
two. We start this section with a quick review of the spectral action 
principle.

\subsection*{The Spectral Action Principle}

The Connes--Chamseddine spectral action principle \cite{Spectral action}
provides a geometric framework in which both gravitational and gauge
interactions arise from the spectral properties of a single operator: the Dirac
operator of a real spectral triple. As given more detail in previous sections, recall that a real even spectral triple consists of
\[
(\mathcal{A},\mathcal{H},D;J,\gamma),
\]
where $\mathcal{A}$ is a $*$-algebra represented on the Hilbert space
$\mathcal{H}$, $D$ is a self-adjoint operator with compact resolvent, $J$ is
the real structure (charge-conjugation operator), and $\gamma$ is the grading.
These data satisfy a set of compatibility axioms determined by the
KO-dimension.

The bosonic action associated to a spectral triple is defined by
\[
S_{\mathrm{bos}}(D,\Lambda)
= \operatorname{Tr}\!\left(f(D/\Lambda)\right),
\]
where $f$ is a positive even cutoff function and $\Lambda$ is a physical energy
scale.  Fermionic matter fields are incorporated through
\[
S_{\mathrm{ferm}}(\psi,D)=\langle\psi, D \psi\rangle,
\qquad \psi\in \mathcal{H}.
\]

A key ingredient of the framework is the notion of \emph{inner fluctuations} of
the Dirac operator:
\[
D ~\longmapsto~ D_{A}=D + A + JAJ^{-1},
\qquad 
A=\sum a_{i}[D,b_{i}],\qquad a_{i},b_{i}\in\mathcal{A},
\]
which play the role of gauge and Higgs fields.
In this viewpoint, bosonic fields arise not as independent dynamical
variables but as generalized connections on the noncommutative space.

\medskip

A canonical example is afforded by a Riemanian spin manifold. 
For a compact Riemannian spin manifold $(M,g)$, the associated canonical
spectral triple is
\[
\mathcal{A}=C^{\infty}(M),\, \, 
\mathcal{H}=L^{2}(M,S), \, \, 
D=D_{M},
\]
where $S$ is the spinor bundle.
Connes’ distance formula reconstructs the geodesic distance from~$D$,
demonstrating that the Dirac operator encodes the full Riemannian geometry.

\medskip

A noncommutative example is the so called almost commutative geometries.  
To describe internal symmetries and particle content, one considers the
tensor product of a manifold with a finite noncommutative space:
\[
\mathcal{A}=C^{\infty}(M)\otimes\mathcal{A}_{F},\qquad  
\mathcal{H}=L^{2}(M,S)\otimes H_{F},\qquad
D=D_{M}\otimes1+\gamma_{5}\otimes D_{F}.
\]
Here $\mathcal{A}_{F}$ is a finite-dimensional algebra and
$(H_{F},D_{F})$
encodes internal degrees of freedom such as Yukawa couplings,
fermion mixing matrices (CKM and PMNS), and Majorana mass terms.
Inner fluctuations of $D$ now generate simultaneously:
\begin{itemize}
\item gauge potentials $A_{\mu}$ associated to the unitary group of $\mathcal{A}_{F}$,
\item the Higgs field, which appears as the finite part of the fluctuation.
\end{itemize}
This structure underlies the reconstruction of the Standard Model, treated in
the next subsection.

\medskip

Let us next discuss the heat-kernel expansion and emergence of physics. 
The bosonic action is evaluated using the heat-kernel asymptotics of $D^{2}$.
For large $\Lambda$,
\[
\operatorname{Tr}(f(D/\Lambda))
~\sim~
\sum_{k\ge 0} f_{k}\,\Lambda^{4-k}\,a_{k}(D^{2}),
\]
where the coefficients $a_{k}(D^{2})$ are the Seeley--DeWitt invariants.
In four dimensions, one finds the hierarchy
\begin{itemize}
\item $\Lambda^{4}$ term: cosmological constant,
\item $\Lambda^{2}$ term: Einstein--Hilbert action and scalar curvature,
\item $\Lambda^{0}$ term: Yang--Mills kinetic terms and the Higgs kinetic term,
\item $\Lambda^{-2}$ and lower: higher-curvature corrections.
\end{itemize}

The fermionic term,
\[
S_{\mathrm{ferm}}(\psi,D_{A})
= \langle \psi, D_{A}\psi\rangle,
\]
contains the usual Dirac kinetic terms, the Yukawa couplings determined by the
finite Dirac operator $D_{F}$, and the Majorana masses  permitted by the
KO-dimension.

\medskip

In the spectral action approach the Dirac operator is a fixed geometric object:
its dynamics do not arise from a functional integral over the space of
operators, but from the asymptotic expansion of a spectral functional.
Thus the spectral action is a \emph{semiclassical} theory: it produces
classical gravitational and gauge actions unified at a geometric level, but
does not provide a quantization scheme for the full spectral data.
Quantization issues and the status of gravity in this framework will be
addressed later, after the reconstruction of the Standard Model.

\subsection{Reconstruction of the Standard Model}

Almost-commutative geometries of the form
\[
\mathcal{A}=C^{\infty}(M)\otimes \mathcal{A}_{F}, \qquad
\mathcal{H}=L^{2}(M,S)\otimes H_{F}, \qquad
D = D_{M}\otimes 1+\gamma_{5}\otimes D_{F},
\]
provide the geometric setting for the spectral reconstruction of the Standard
Model.  Here $M$ is a compact four-dimensional Riemannian spin manifold and
$\mathcal{A}_{F}$ is the internal algebra
\[
\mathcal{A}_{F}
  = \mathbb{C} \oplus \mathbb{H} \oplus M_{3}(\mathbb{C}),
\]
whose unitary group reproduces, up to a finite quotient, the gauge group
\[
U(1)_{Y}\times SU(2)_{L}\times SU(3)_{c}.
\]
The Hilbert space $H_{F}$ encodes the fermionic content of one generation, and
the finite Dirac operator $D_{F}$ contains the Yukawa couplings, Majorana mass
terms, and inter-generational mixing matrices.  The inner fluctuations of the
Dirac operator,
\[
D \mapsto D_{A}=D+A+JAJ^{-1},
\]
generate both gauge and scalar fields.  In this way the fluctuated operator
produces the gauge bosons associated with $U(1)_{Y}$, $SU(2)_{L}$ and
$SU(3)_{c}$, together with the Higgs field arising from the off-diagonal part of
the finite fluctuation.  Nothing is introduced by hand: the full field content
of the Standard Model emerges from the geometry.

The bosonic spectral action,
\[
S_{\mathrm{bos}}(D,\Lambda)=\operatorname{Tr}\bigl(f(D/\Lambda)\bigr),
\]
admits an asymptotic heat-kernel expansion that reproduces the complete
Lagrangian of the Standard Model coupled to gravity.  The expansion yields the
Einstein--Hilbert term with cosmological constant, the Yang--Mills actions for
the gauge fields, and the entire Higgs sector including kinetic, mass, and
quartic terms.  Fermion masses and Yukawa couplings are encoded in the
eigenvalues of $D_{F}$, and the presence of right-handed neutrinos in $H_{F}$
leads naturally to the seesaw mechanism.  A geometric unification relation
follows:
\[
g_{3}^{2}=g_{2}^{2}=\frac{5}{3}g_{1}^{2}
\qquad \text{at the unification scale},
\]
capturing the origin of the gauge couplings from the internal algebra.

On the gravitational side, the expansion
\[
S_{\mathrm{bos}}(D_{M},\Lambda)
=\Lambda^{4}f_{4}a_{0}
+ \Lambda^{2}f_{2}a_{2}
+ f_{0}a_{4}
+ \cdots
\]
shows that the coefficient $a_{2}$ contains the scalar curvature and reproduces
the Einstein--Hilbert action.  The higher coefficients generate
curvature-squared terms, which become important in the ultraviolet regime and
play a significant role in early-universe cosmology.

The Higgs potential also arises geometrically.  The $\Lambda^{0}$ term of the
expansion produces the quartic potential
\[
V(H)=\lambda |H|^{4}-\mu^{2}|H|^{2},
\]
where the coefficients are determined by traces of powers of $D_{F}$.  This
provides a conceptual shift: the Higgs potential is not inserted into the
Lagrangian but derived from the internal geometry of the spectral triple.

In the two-sheet structure underlying the finite noncommutative space, the
fluctuated Dirac operator takes the block form
\[
D_{A}=
\begin{pmatrix}
D_{M}+A_{\mu}\gamma^{\mu} & \Phi + H \\
(\Phi + H)^{\ast} & D_{M}+B_{\mu}\gamma^{\mu}
\end{pmatrix},
\]
where $A_{\mu}$ and $B_{\mu}$ are gauge fields on each sheet, and $H$ is a
complex scalar field.  The diagonal components correspond to the Standard Model
gauge bosons, while the off-diagonal component yields a Higgs-type field.
Geometrically, the Higgs field measures the failure of the two sheets to remain
decoupled, and the distance in the discrete direction of the internal space is
inversely proportional to $|\Phi|$.  The spectral action automatically produces
a potential of the form
\[
V(\Phi)=\lambda |\Phi|^{4}-\mu^{2}|\Phi|^{2},
\]
leading to spontaneous symmetry breaking.  Thus the Higgs mechanism is the
manifestation of the finite noncommutative geometry encoded in $(\mathcal{A}_{F},H_{F},D_{F})$.

Higher-order terms in the spectral action expansion introduce curvature-squared
corrections that have been shown to give rise to viable inflationary potentials
\cite{Marcolli2012, Marcolli2017, Marcolli2019}.  This demonstrates that the
same geometric framework that reconstructs particle physics and gravity also
yields a natural setting for early-universe cosmology.

Finally, while the spectral action provides a remarkably unified classical
description of matter and geometry, the Dirac operator remains a fixed
geometric object, and a full theory of quantum fluctuations of $D$ compatible
with the axioms of noncommutative geometry is still lacking.  Recent work on
random Dirac operators and ensembles of finite spectral triples suggests a
possible route toward quantizing geometry by replacing a single Dirac operator
with a probability measure on the space of operators.  How such ensembles can
incorporate the almost-commutative structure $\mathcal{A}_{F}$ and reproduce
the Standard Model sector remains an intriguing and promising direction.

\subsection{Towards a Noncommutative Connes--Lott Model}

The Connes--Lott model \cite{Connes-Lott, Connes94, QFTNCG} provides the earliest and simplest realization of the idea that the internal symmetries of particle physics arise as noncommutative directions of space-time.  It serves as a conceptual and mathematical precursor to the full Chamseddine--Connes spectral action formulation of the Standard Model.  In particular, it offers a geometric justification for coupling a fluctuating Dirac operator---or, in the probabilistic setting, a Dirac ensemble---to the finite noncommutative space encoding the internal degrees of freedom.

The underlying noncommutative space is the product
\[
X = M \times \{0,1\},
\]
where $M$ is a smooth compact four-dimensional spin manifold (space-time), and $\{0,1\}$ is a discrete two-point space.  The algebra of ``functions'' on $X$ is
\[
\mathcal{A} = C^{\infty}(M) \otimes (\mathbb{C}\oplus\mathbb{C})
           \cong C^{\infty}(M) \oplus C^{\infty}(M),
\]
whose elements have the form
\[
a(x)=
\begin{pmatrix}
f(x) & 0 \\
0 & g(x)
\end{pmatrix},
\qquad f,g\in C^{\infty}(M).
\]

The Hilbert space is the direct sum of two copies of the spinor space,
\[
\mathcal{H}=L^{2}(M,S)\oplus L^{2}(M,S).
\]
The Dirac operator on $X$ takes the form
\[
D =
\begin{pmatrix}
D_{M} & \Phi \\
\Phi^{\ast} & D_{M}
\end{pmatrix},
\]
where $D_{M}$ is the usual Dirac operator on $M$, and $\Phi\colon L^{2}(M,S)\to L^{2}(M,S)$ is a scalar field on~$M$.  In the full noncommutative Standard Model, $\Phi$ becomes the Higgs field.
A straightforward computation shows
\[
[D,a] =
\begin{pmatrix}
[D_{M},f] & \Phi g - f\Phi \\
\Phi^{\ast} f - g\Phi^{\ast} & [D_{M},g]
\end{pmatrix}.
\]
The off-diagonal components measure the interaction between the two sheets: they represent a parallel transport in the discrete direction and therefore encode a scalar field.  This is the geometric origin of the Higgs field in the Connes--Lott model.

\medskip

The Connes--Lott picture shows in a simple but concrete setting how internal symmetries can be understood as noncommutative directions of space-time, and how scalar fields such as the Higgs arise from the geometry of a finite spectral triple.  This provides the natural bridge to the Dirac ensemble approach, where the finite Dirac operator is allowed to fluctuate, and the finite noncommutative geometry becomes a dynamical—indeed probabilistic—internal space. In particular, a key insight of the Connes--Lott construction is that the finite Dirac operator~$\Phi$ is not fixed by symmetry and thus carries physical information such as Yukawa couplings.  There is no geometric or physical reason for $\Phi$ to remain classical.  This observation motivates the Barrett--Glaser program of integrating over the space of all finite Dirac operators,
\[
Z \;=\; \int \exp\!\bigl(-S(D)\bigr)\, dD,
\]
leading naturally to the notion of a \emph{Dirac ensemble}: a probability measure on the space of finite Dirac operators that describes quantum fluctuations of the internal geometry.  Since the Higgs field and Yukawa couplings appear as components of~$\Phi$, integrating over~$\Phi$ amounts to integrating over an extended class of geometries.  The Connes--Lott model thus provides a simple but meaningful geometric example illustrating why such an approach is natural.

\subsection{Building a Bridge}

The finite noncommutative space appearing in the full Chamseddine--Connes model has the
algebra
\[
\mathcal{A}_{F} = \mathbb{C} \oplus \mathbb{H} \oplus M_{3}(\mathbb{C}),
\]
together with a finite-dimensional Hilbert space carrying the fermionic degrees of freedom
and a finite Dirac operator whose components encode Yukawa couplings, mixing angles, and
mass matrices.  
Although this finite geometry is considerably more intricate than the two-point example,
the conceptual mechanism remains identical: the off-diagonal entries of the finite Dirac
operator give rise to the Higgs field, the diagonal entries determine fermion masses,
gauge fields arise through inner fluctuations, and the bosonic sector is obtained from the
spectral action.

In this sense, the Connes--Lott model is the simplest prototype of the finite geometry
underlying the Standard Model.  
The idea that the Higgs field appears as a connection in a discrete internal direction
survives unchanged in the full theory.

A different perspective is provided by the Dirac ensemble approach, in which the Dirac
operator itself becomes a fluctuating variable.  
Starting from a set of fuzzy geometries, one considers a statistical partition function
\[
Z = \int_{\mathcal{D}} \exp\!\left( - S(D) \right)\, dD,
\]
where $\mathcal{D}$ is the space of all Dirac operators compatible with the given
finite spectral triple structure.  
The action is typically chosen to be a polynomial so that the model becomes a multi-matrix
integral in which different Dirac operators represent different geometries.  
The partition function thus defines a probability measure on the space of geometries,
yielding a genuine quantization of finite noncommutative spaces.  
Large-$N$ limits, universality properties, and phase transitions can be analyzed using
methods from random matrix theory.

\medskip

From a conceptual standpoint, the two frameworks differ in significant ways.
In the spectral action approach, geometry is fixed and enters as a background structure,
whereas in the Dirac ensemble picture geometry becomes dynamical because one integrates
over the space of Dirac operators.  
The spectral action is defined through the nonlocal spectral functional
$\mathrm{tr}(f(D/\Lambda))$, whose asymptotic expansion produces curvature invariants, 
while Dirac ensembles employ polynomial potentials resembling matrix-model actions.  
The geometric settings also diverge: the spectral action lives on smooth or
almost-commutative manifolds, whereas Dirac ensembles begin with fuzzy geometries in
which the Dirac operator encodes all geometric information.  
Gauge fields arise from inner fluctuations in the spectral action, while in the Dirac
ensemble framework no separate gauge sector is introduced—everything is contained within
the fluctuating Dirac operator.

The mathematical techniques used in each setting reflect the same contrast.
The spectral action relies on heat-kernel expansions, Seeley--DeWitt coefficients, and
analytic tools from spectral geometry, whereas Dirac ensembles employ Schwinger--Dyson
equations, Coulomb-gas methods, topological recursion, and bootstrap techniques.

A similar distinction appears in their physical interpretations.
The spectral action provides a classical unified Lagrangian for gravity and the Standard
Model, with all bosonic fields emerging from geometry.  
Dirac ensembles instead offer a statistical, non-perturbative model of quantum geometry in
which the Dirac operator fluctuates and different geometries contribute probabilistically.
The two perspectives are therefore complementary: one geometric and deterministic, the
other quantum and statistical.

\medskip

Regarding spectral geometric properties, the heat kernel expansion is the basic tool for
analyzing the Dirac operator of spectral triples.  
For a Dirac operator on a manifold of dimension $d$, one has the asymptotic expansion
\begin{equation*}
\sum_{\lambda \in \text{spec}(D)} e^{-t\lambda^2}
   \sim_{t \rightarrow 0} t^{-d/2}(a_{0}+a_{2}t + a_{4}t^2),
\end{equation*}
where the coefficients $a_i$ encode geometric invariants such as the volume and the
Riemann curvature~\cite{Gilkey}.  
This allows one to associate classical geometric quantities to spectral triples.

In contrast, the spectrum of Dirac operators in fuzzy settings typically decays
super-exponentially, so such an asymptotic expansion is unavailable.  
In the large-$N$ limit the spectrum is often compact.  
Instead one considers the normalized heat kernel
\begin{equation*}
K_{D^2}(t):=\lim_{N \rightarrow \infty}\frac{1}{Z}\mathbb{E}\!\left[\tr e^{-t D^2}\right],
\end{equation*}
from which one may compute the spectral dimension
\begin{equation*}
d_{s}(t) := -2t\frac{d}{dt}\log(K_{D^2}(t)),
\end{equation*}
and the spectral variance
\begin{equation*}
v_{s}(t) := 2t^2\frac{d^{2}}{dt^{2}}\log(K_{D^2}(t)).
\end{equation*}
Both quantities were introduced in~\cite{Spectral estimators} and relate to the metric
dimension in deterministic fuzzy geometries.  
As noted in~\cite{Spectral estimators}, the development of notions such as random volume and
random curvature would be very interesting.

\medskip

One may also imagine combining the two frameworks.  
Formally inserting the spectral action into a Dirac ensemble,
\[
Z = \int_{\mathcal{D}} \exp\!\left( - \mathrm{tr}(f(D/\Lambda)) \right)\, dD,
\]
would produce a fully non-perturbative model of geometry and matter.  
Such a theory remains far beyond current analytic control, but the idea hints at a deep
structural connection between spectral geometry and probabilistic models of
noncommutative spaces.

\medskip

\begin{center}
\begin{tabular}{lll}
\hline
\textbf{Feature} & \textbf{Spectral Action} & \textbf{Dirac Ensembles} \\
\hline
Geometry & fixed background & fluctuating \\
Dirac operator & geometric input & random variable \\
Action & $\mathrm{tr}(f(D/\Lambda))$ & polynomial in $D$ \\
Physics & gravity + SM & quantum geometry \\
Methods & heat kernel & random matrices \\
Goal & classical unification & nonperturbative quantization \\
\hline
\end{tabular}
\end{center}

The spectral action provides a unified geometric action for known physics, while Dirac
ensembles offer a statistical description of fluctuating noncommutative geometries.
Together, they illustrate two complementary ways in which noncommutative geometry may
illuminate the structure of fundamental interactions.

\subsection{Fermionic Dirac Ensembles}
\label{sec:summary-fuzzy-fermionic}

Adding fermions to path integrals over finite spectral triples was explored in generality setting in \cite{Barrett2024}. In this section we focus on and summarize the main results of
Khalkhali--Pagliaroli--Verhoeven~\cite{Khalkhali2025 large N limit},
which develops a nonperturbative large-$N$ theory for quartic fuzzy Dirac
ensembles coupled to fermions.  
These ensembles arise from $(0,1)$ fuzzy spectral triples and constitute
single-matrix multi-trace random matrix models whose spectral properties 
encode the geometry of the underlying finite noncommutative space.  
The paper establishes the existence of a large-$N$ spectral density for a 
broad class of multi-trace, non-polynomial matrix models and studies 
properties such as spectral dimension, spectral variance, and 
phase transitions. In theory, this framework for adding fermions with mass works for other signatures, but our attention is restricted to analytically tractable models.

\subsubsection{Fuzzy Geometries and $(0,1)$ Spectral Triples}

For type $(0,1)$ fuzzy geometries, the Clifford module is one-dimensional and the Dirac 
operator takes the form
\[
D = [H,\cdot], \qquad H = H^\ast \in M_{N}(\mathbb{C}).
\]
Thus the spectrum is given by the eigenvalue differences
\begin{equation}
	\label{eq:spectrum-dfuzzy-plain}
	\mathrm{spec}(D)
	=
	\{\lambda_i - \lambda_j : 1\le i,j\le N\},
\end{equation}
each appearing with multiplicity one.  
This doubling phenomenon reflects a familiar feature of finite spectral 
triples \cite{Connes94, QFTNCG}. The space of Dirac operators of the $(0,1)$ fuzzy geometry thus reduces to
\[
D_{N} = \{[H,\cdot] : H \in M_{N}(\mathbb{C})_{\text{Herm}}\},
\]
so Dirac ensembles correspond to random matrix ensembles for the matrix $H$. Because $D=[H,\cdot]$ is invariant under $H\mapsto H+cI$, to define a Dirac ensemble one introduces a 
Gaussian regularization on the trace part of $H$,
\[
\exp(-a\, \mathrm{Tr}(H)^{2}),
\]
yielding an absolutely continuous measure on the space of Hermitian matrices.

Pulling back the ensemble along $H\mapsto D=[H,\cdot]$ produces a 
multi-trace model of the form
\[
p(H) \propto 
\exp\!\left(
-S(D) - a\, \mathrm{Tr}(H)^{2}
\right),
\]
where $D_{H} = [H,\cdot]$.  
Since
\[
D_{H}^{k}
=
(H\otimes 1 - 1\otimes H)^{k},
\]
the action contains traces of powers of $H\otimes 1 - 1\otimes H$.

We take the complex Grassmann algebra generated by our Hilbert space of complex $N$ by $N$ matrices as our space of fermions. As discussed in \cite{Barrett2024} real or quaternionic algebras are also very possible, but for simplicity of presentation we restrict ourselves to the complex case. In this case the partition function of our Dirac ensemble is of the form
\begin{equation*}
	\int_{\mathcal{H}_{N}}e^{-S(D) -\langle \psi, D \psi \rangle}dHd \psi = 	\int_{\mathcal{H}_{N}}\det(D)e^{-S(D)} dH.
\end{equation*}
By first integrating out the fermionic action, we can define an honest probability distribution on $D$ and use techniques from random matrix theory to compute random determinants. However, since $D=[H,\cdot]$ has a large kernel, $\det(D)=0$ identically.

To obtain a nontrivial determinant, one couples the fuzzy spectral triple to a 
finite spectral triple describing a single massive fermion as is done in the noncommutative geometric approach to particle physics \cite{VS}. We accomplish this by considering models that come from type $(0,1)$ fuzzy geometries with a finite spectral triple
\begin{equation*}
	(\mathbb{C},\mathbb{C},m;J=\bar{\cdot})
\end{equation*}
with $KO$-dimension seven. The resulting product spectral triple is
\begin{equation}
	\left(M_N(\mathbb{C}), M_N(\mathbb{C}) \otimes \mathbb{C}^2, D=D_{f u z z y} \otimes \sigma_1+m \otimes \sigma_2 ; J, \Gamma=1 \otimes \sigma_3\right),
\end{equation}
with $KO$ dimension zero and where each $\sigma_{i}$ is a Pauli matrix. Note that naively adding a mass term to the Dirac operator i.e. $D+m$, does not satisfy the requirement of a type $(0,1)$ fuzzy geometry that $JD = -DJ$. Its spectrum of $D = D_{\mathrm{fuzzy}} \otimes \sigma_{1} 
+ m \otimes \sigma_{2}$ becomes
\begin{equation}
	\label{eq:massive-spectrum-plain}
	\mathrm{spec}(D)
	=
	\left\{
	\pm\sqrt{m^{2} + (\lambda_i - \lambda_j)^{2}} 
	\;\middle|\; 1\le i,j\le N
	\right\}.
\end{equation}
Hence, integrating out the fermions yields the effective contribution to the action of
\begin{equation}
	\label{eq:fermionic-log-plain}
	\sum_{i,j} 
	\log\!\bigl(m^{2}+(\lambda_{i}-\lambda_{j})^{2}\bigr),
\end{equation}
which produces eigenvalue repulsion.

Note that if we wanted to consider areal Grassmann algebra, in action would replace $\langle \psi, D\psi\rangle$ with $\frac{1}{2}\langle J \psi, D \psi\rangle$.

\subsection{The Quartic Model}
The main theoretical result of \cite{Khalkhali2025 large N limit} is a proof of the existence and
uniqueness of the eigenvalue density function in the large $N$ limit. Under suitable convexity, symmetry, and growth hypotheses on the choice of $S(D)$, for a given coupling constant configuration,
there exists a unique compactly supported measure such that
\[
\lim_{N\to\infty}
\frac{1}{N}\mathrm{Tr}(f(H))
=
\int f(x)\, d\mu_{E}(x)
\]
for all tracial observables $f$.
This result extends classical logarithmic potential theory for single-trace
models to multi-trace, non-polynomial interactions. 

The quartic  Dirac ensemble is defined by the probability density
\begin{equation}
	\label{eq:bosonic-action-plain}
	p(D) = \frac{1}{Z_{D, \psi}} \exp \left(-\operatorname{Tr}\left(g_4 D^4+g_2 D^2\right)+\frac{\beta_2}{4} \sum_{i, j} \log \left(m^2+\left(\lambda_i-\lambda_j\right)^2\right)\right) ,
\end{equation}
with $g_{4}>0$ to ensure convergence.  Applying Weyl's integration formula gives the joint eigenvalue distribution density function as 
\begin{align*}
	p(\lambda)
	&\propto \exp\!\left(
	-2g_{4}\!\sum_{i,j}(m^{2}+(\lambda_{i}-\lambda_{j})^{2})^{2}
	-2g_{2}\!\sum_{i,j}(m^{2}+(\lambda_{i}-\lambda_{j})^{2})
	+\frac{1}{2}\!\sum_{i,j}\log(m^{2}+(\lambda_{i}-\lambda_{j})^{2})\right.\\
	&\left. +\frac{1}{2}\!\sum_{i\neq j}\log(\lambda_{i}-\lambda_{j})^{2}
	-a\Bigl(\sum_{i}\lambda_{i}\Bigr)^{2}
	\right),
\end{align*}
a highly non-polynomial multi-trace ensemble.

\begin{figure}[H]
 	\centering
 	\includegraphics[width=\textwidth]{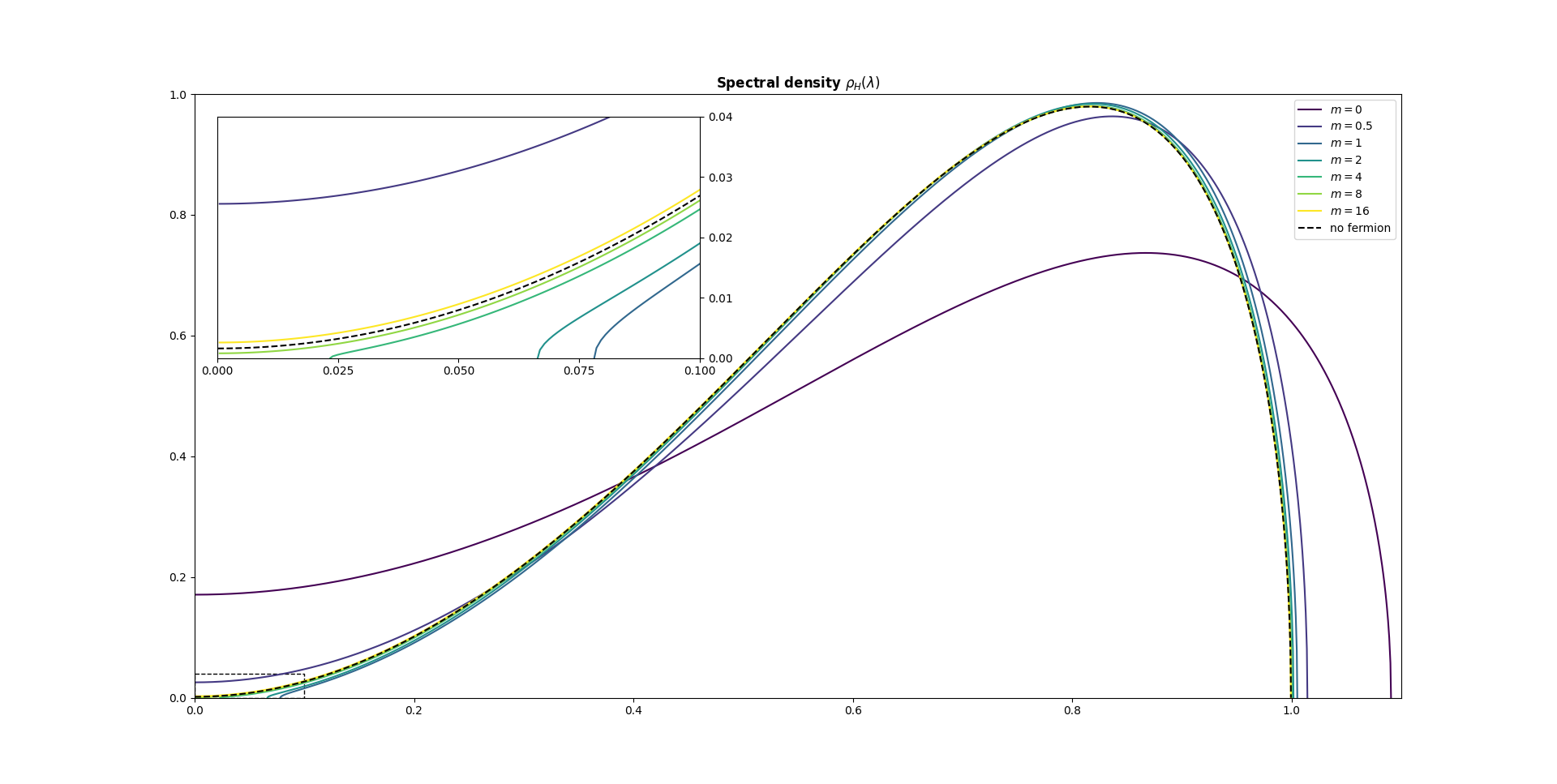}
 	\caption{The positive part of the spectral density of $H$ for the fixed couplings $g_2 = -3.99$, $g_4 = 1$, for various values of $m$. Observe that for $m = 0$, $m = 0.5$, $m= 8$, and $m = 16$ the model is in the one-cut phase, while for $m = 1$, $m=2$, and $m=4$ the model is in the two-cut phase. In these cases the repulsive effect of the fermions being is stronger between the two wells than within one well well if the mass is approximately the separation between the wells \cite{Khalkhali2025 large N limit}.}
 	\label{fig:anomalous_force}
 \end{figure}

 The quartic type $(0,1)$ model indeed satisfies the conditions for uniqueness  and the probability density function of the ensemble in the large $N$ limit can be solved for a s the solution of a complicated integral equation:
 \begin{equation*}
 	\text { P. V. } 
 	\int_{\Sigma} \frac{\rho(y)}{y-x} d y  =- \frac{d}{d x} \int_{\Sigma}\left(2 g_4(x-y)^4+2 g_2(x-y)^2-\frac{1}{2} \log \left(m^2+(x-y)^2\right)+\frac{a}{2} x y\right) \rho(y) d y 
 \end{equation*}
 The equation can be solved numerically, see Figure \ref{fig:anomalous_force}. One can see that just like in the non-fermionic case, there is a spectral phase transition where the support of the eigenvalue distribution splits into two intervals for $H$. 
 
	\begin{figure}[H]
	\centering
	\includegraphics[width=\textwidth]{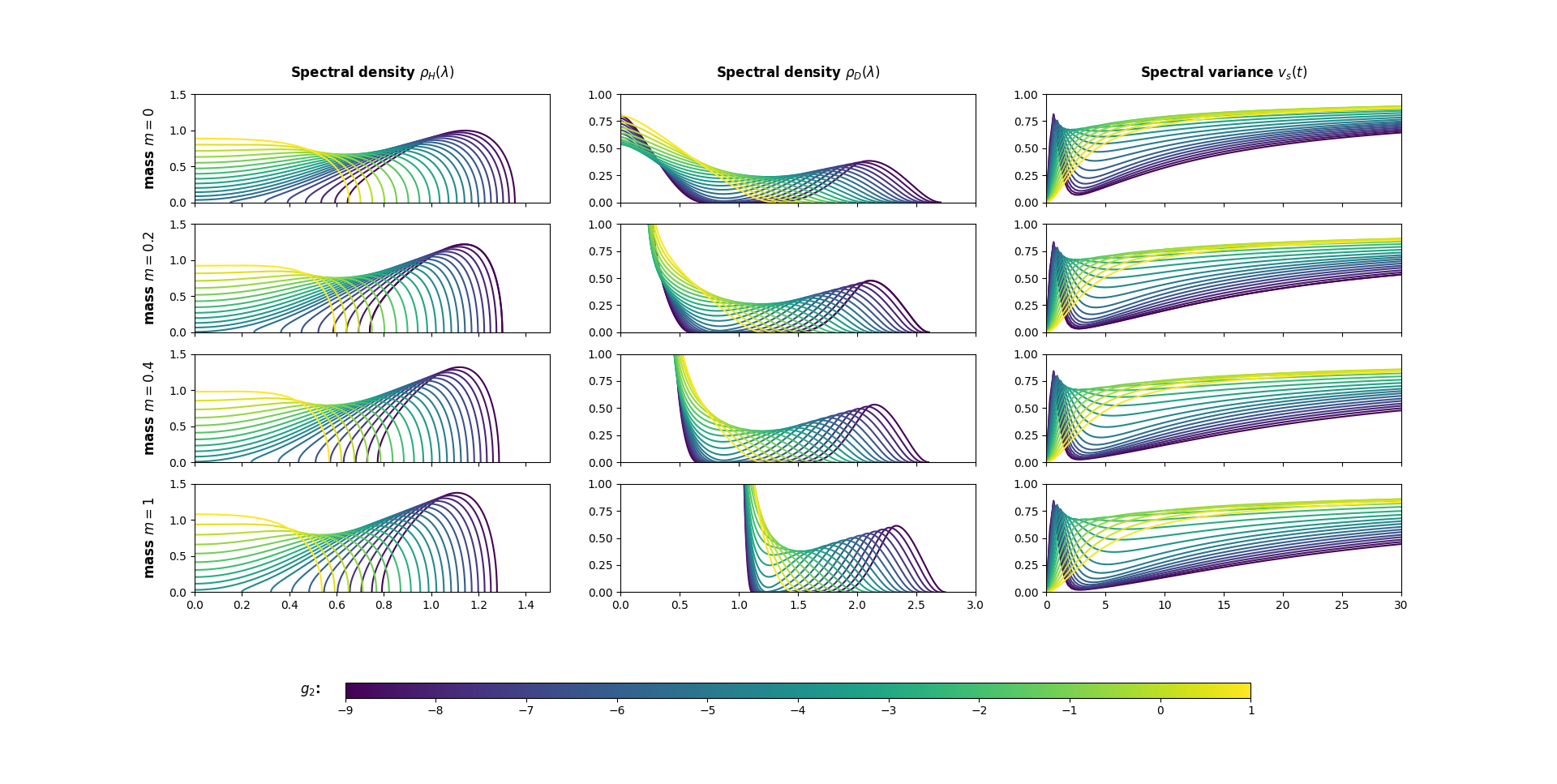}
	\caption{For the type $(0,1)$ fermionic Dirac ensemble, the distribution of eigenvalues of $H$ and $D$, as well as the spectral variance for various values of $g_2$ and masses. The coupling $g_4$ is fixed at one. Negative values of $g_2'$ (darker color) correspond to two-cut models, when the double well becomes more pronounced. The spectral gap for the eigenvalue distribution of the Dirac operator increases, which corresponds to the dip in the spectral variance, while the spacing between the phases increases.}
	\label{fig:influence_of_g2}
\end{figure}

Fermionic Dirac ensembles provide a promising bridge between
noncommutative geometry and random matrix theory.  
They suggest a pathway toward a quantized version of the spectral action,
in which both the metric and finite internal geometry fluctuate within 
a matrix ensemble framework.  
Coupling to richer finite spectral triples—such as those of the 
Connes--Chamseddine Standard Model—remains an important direction for 
future development.

Recent work of the authors and a collaborator has studied the SDE's of type $(0,1)$ models coupled to fermions as well as bosons \cite{Gamble2026}. The loop equations can be solved recursively. In the Gaussian case of both, the solution can be explicitly written in terms elliptic integrals and the coupling constant. Additionally, the free energy of these models is related to the generating function of three-coloured triangulations of the 2-sphere.

\subsection{Yang--Mills--Higgs Dirac Ensembles}

In a fuzzy geometry, one can think of $V_{p,q}$ as a stand-in for a spinor module and of $M_N(\mathbb{C})$ as the
analogue of square-integrable functions, so that together they form the simplest possible
spinor bundle over a noncommutative space.  
But if the ultimate goal is to build models that resemble the gauge theories of physics,
metric information alone is not enough: the geometry must also be able to carry gauge and
Higgs fields.  

A natural way to achieve this is to enlarge the fuzzy spectral triple by coupling it to a
finite spectral triple, much in the spirit of almost-commutative manifolds
\cite{Spectral action, Neutrino Mixing}.  
In this viewpoint the finite triple plays the role of an internal vector bundle on which a
connection may act.  
This philosophy underlies the Yang--Mills--Higgs fuzzy spaces introduced by Perez-Sanchez in\cite{Sanchez Yang-Mills}, which we will now review.

We start with a fuzzy spectral triple
\[
M_f = (M_N(\mathbb{C}), \mathcal{H}_N, D_f, J, \gamma),
\]
and pair it with a finite spectral triple
\[
F = (\mathcal{A}_F, \mathcal{H}_F, D_F, J_F, \gamma_F).
\]
This second triple carries the additional degrees of freedom that will eventually manifest
themselves as gauge bosons and Higgs fields.  
When $\mathcal{A}_F = M_n(\mathbb{C})$ and $\mathcal{H}_F = M_n(\mathbb{C})$, we obtain what
is known as a Yang--Mills triple.  
The combined system is the product spectral triple
\[
\left(
M_N(\mathbb{C}) \otimes \mathcal{A}_F,\,
\mathcal{H}_N \otimes \mathcal{H}_F,\,
D_f \otimes 1 + \gamma \otimes D_F,\,
J \otimes J_F,\,
\gamma \otimes \gamma_F
\right).
\]
In this description the role of $D_F$ is to carry the gauge-theoretic data, while $D_f$
continues to govern the underlying fuzzy geometry.  
As $N \to \infty$ the fuzzy geometry approaches a smooth space, while the internal dimension
$n$ remains fixed.  
Recalling that
\[
D_f = \sum \gamma^I \otimes \{ K_I, \cdot \}_{e_I},
\]
the amplified operator $D_f \otimes 1$ becomes
\[
D_f \otimes 1 = \sum \gamma^I \otimes \{ K_I \otimes 1_n, \cdot \}_{e_I}
\]
acting naturally on $V_{p,q} \otimes M_N(\mathbb{C}) \otimes M_n(\mathbb{C})$.

\medskip

A key mechanism that brings gauge fields into the picture is the inner fluctuation of the
Dirac operator \cite{Connes Inner Fluctuations}.  
In any spectral triple $(\mathcal{A},\mathcal{H},D,J,\gamma)$, the one-forms
\[
\Omega^1_D(\mathcal{A}) = 
\left\{ \sum a_i[D,b_i] \,\middle|\, a_i,b_i \in \mathcal{A} \right\}
\subset \mathrm{End}(\mathcal{H})
\]
parametrise infinitesimal deformations of $D$.  
A self-adjoint element $\omega \in \Omega^1_D(\mathcal{A})$ produces a new Dirac operator
\[
D_\omega = D + \omega + J\omega J^{-1},
\]
and this fluctuation is the noncommutative analogue of adding a connection to a bundle.

For the product fuzzy Yang--Mills--Higgs triple, one can distinguish two types of
fluctuations: those coming from the fuzzy part and those coming from the finite part.
Indeed, for $a,b \in M_N(\mathbb{C}) \otimes M_n(\mathbb{C})$,
\[
a[D,b] = a[D_f \otimes 1, b] + a[\gamma \otimes D_F, b].
\]
The fuzzy fluctuations sit in
\[
\Omega^1_{D_f}(M_N(\mathbb{C})) \otimes M_n(\mathbb{C}),
\]
and modify the operators $K_I \otimes 1_n$ by adding terms $T_I$ of the correct adjointness.
The finite fluctuations, by contrast, leave the fuzzy sector untouched and produce a new
operator
\[
\Phi = 1_N \otimes D_F + \{\phi, \cdot\}_{\epsilon''},
\qquad 
\phi \in M_N(\mathbb{C}) \otimes \Omega^1_{D_F}(M_n(\mathbb{C})).
\]
This $\Phi$ is the fuzzy incarnation of the Higgs field.

\medskip

To make the geometric meaning more transparent, consider the Riemannian signature
$(p,q)=(0,4)$.  
Here one finds the representation
\[
D_f = \sum_\mu \gamma^\mu \otimes [L_\mu,\cdot]
+ \gamma^{\hat{\mu}} \otimes \{X_\mu,\cdot\}.
\]
The term $[L_\mu,\cdot]$ acts like a discrete partial derivative, while $\{X_\mu,\cdot\}$
resembles a symmetrised connection coefficient.  
When all $X_\mu$ vanish the geometry is naturally interpreted as flat.

The gauge group of a spectral triple $(\mathcal{A},\mathcal{H},D,J,\gamma)$ is
\[
G(\mathcal{A},J) = \{ uJuJ^{-1} \mid u \in U(\mathcal{A}) \}.
\]
For the Yang--Mills fuzzy triple this becomes $PU(N)\times PU(n)$, where the first factor
acts on the fuzzy space and the second acts on the finite part, just as a classical
Yang--Mills gauge group would act on an internal vector bundle.

In the flat $(0,4)$ case the self-adjoint fluctuations may be written as
\[
\sum_\mu \gamma^\mu \otimes A_\mu,
\quad
A_\mu \in \Omega^1_{L_\mu}(M_N(\mathbb{C})) \otimes M_n(\mathbb{C}),
\]
and they modify $L_\mu$ via $L_\mu \mapsto L_\mu + A_\mu$.  
This is precisely the behaviour expected of a gauge connection.  
The corresponding field strength is defined by
\[
F_{\mu\nu}
=
\big[\,[L_\mu + A_\mu,\cdot],\, [L_\nu + A_\nu,\cdot] \big],
\]
mirroring the usual commutative expression.

\medskip

If one considers the quartic potential
\[
S(D) = g\,\mathrm{tr}(D^2) + \mathrm{tr}(D^4),
\]
then for a fluctuated Dirac operator in the $(0,4)$ case one finds \cite{Sanchez Yang-Mills}
\[
S(D) =
-2\tr(F_{\mu\nu}F^{\mu\nu})
+ 4\tr(g\theta + \theta^2)
+ 4\tr(g\Phi^2 + \Phi^4)
- 8\tr([L_\mu+A_\mu,\Phi][L^\mu+A^\mu,\Phi]),
\]
with
\[
\theta = \eta^{\mu\nu}[L_\mu + A_\mu,\cdot]\circ [L_\nu + A_\nu,\cdot].
\]
Each of these contributions parallels a familiar term in classical Yang--Mills--Higgs
theory:
\begin{itemize}
\item $\tr(F_{\mu\nu}F^{\mu\nu})$ corresponds to the Yang--Mills action,
\item $\tr(g\theta + \theta^2)$ reflects geometric information in the spirit of a Laplacian and its heat-kernel expansion,
\item $\tr(g\Phi^2 + \Phi^4)$ yields the Higgs potential,
\item $\tr([L_\mu+A_\mu,\Phi][L^\mu+A^\mu,\Phi])$ represents the gauge--Higgs coupling.
\end{itemize}

\medskip

Combining a fuzzy spectral triple with a finite one and allowing inner fluctuations thus
provides a natural entry point for Yang--Mills and Higgs physics in the Dirac ensemble
framework.  
In the $(0,4)$ case, the space of Dirac operators $\mathcal{D}$ contains the skew-adjoint
matrices $L_\mu$, their inner fluctuations
$A_\mu \in \Omega^1_{L_\mu}(M_N(\mathbb{C}))$, and the Higgs-type elements
$\phi \in M_N(\mathbb{C}) \otimes \Omega^1_{D_F}(M_n(\mathbb{C}))$, while $D_F$ itself is
kept fixed.  
These variables are not freely independent---the $A_\mu$ arise from a single fuzzy one-form
$\omega \in \Omega^1_{D_f}(M_N(\mathbb{C}))$—which complicates any analytic attempts to study
the ensemble.

Nevertheless, numerical explorations of Yang--Mills--Higgs Dirac ensembles, especially in
lower-dimensional fuzzy settings, appear highly promising.  
Another natural direction would be the inclusion of fermions in the ensemble, for instance
through an additional term of the form $\langle D\psi, \psi \rangle$ with 
$\psi \in \mathcal{H}_N \otimes \mathcal{H}_F$, thereby integrating fermionic and geometric
degrees of freedom into a single probabilistic framework.

\section{Bootstrap Methods in Spectral Geometry}

\subsection{The Bootstrap Philosophy in the Geometric Setting}

%The remarkable feature of the bootstrap program is that no dynamical input is required. That is,  the theory is constrained, and sometimes uniquely determined, by symmetry and positivity alone.

As we already briefly explained  above,  the bootstrap philosophy originated in the study of conformal field theory (CFT),
asserts that a theory may be determined not from a Lagrangian or geometric model,
but from a set of algebraic consistency conditions.  In CFT, these conditions take
the form of the \emph{crossing symmetry} of four-point correlation functions together
with \emph{unitarity} (positivity of norms), leading to nonlinear constraints on
operator dimensions and OPE coefficients \cite{BPZ1984,RRTV}.  In the geometric setting, an analogous structure emerges when studying Laplace
eigenfunctions on a compact Riemannian manifold $(M,g)$, or more generally on
a quotient of a Lie group by a lattice $\Gamma \subset G$.  The role of
CFT correlation functions is played by integrated products of eigenfunctions,
such as
\[
\int_M \varphi_i(x)\, \varphi_j(x)\, \varphi_k(x)\, \varphi_\ell(x)\, d\mathrm{vol}(x),
\]
or, in the automorphic setting,
\[
\int_{\Gamma \backslash G} F_1(g)\, F_2(g)\, F_3(g)\, F_4(g)\, dg.
\]
These objects admit two algebraically distinct decompositions (e.g.\ into
triple product coefficients or spectral channels), and equating them produces
\emph{bootstrap identities} formally parallel to the CFT crossing equations.
The Laplace eigenvalues and triple product coefficients play the role of operator
dimensions and OPE coefficients, and the positivity of the $L^2$ inner product
supplies the analogue of unitarity.  

%Recent work has begun to make this analogy systematic, particularly for hyperbolic manifolds and Einstein spaces \cite{Bon1,Bon2}.

\medskip

Concretely, if $\{\varphi_k\}$ is an orthonormal basis of Laplace eigenfunctions
with eigenvalues $\lambda_k$, one introduces the coefficients
\[
C_{ijk} := \int_M \varphi_i(x)\, \varphi_j(x)\, \varphi_k(x)\, d\mathrm{vol}(x),
\]
and the spectral four-point function
\[
\mathcal{G}_{ij,kl}
:= \int_M \varphi_i(x)\, \varphi_j(x)\, \varphi_k(x)\, \varphi_\ell(x)\, d\mathrm{vol}(x).
\]
Using completeness of the eigenbasis yields the decomposition
\[
\mathcal{G}_{ij,kl}
= \sum_{m} \frac{C_{ijm} \, C_{klm}}{\lambda_m - \lambda_0},
\]
while permutation symmetry (e.g.\ $j \leftrightarrow k$) gives a second expansion.
Equating the two yields an infinite family of algebraic relations among
$\{\lambda_k\}$ and $\{C_{ijk}\}$, which serve as \emph{bootstrap constraints}.

\medskip

Positivity enters through Gram matrices of the form
\[
\mathcal{M}_{ij} = \int_M f_i(x)\, f_j(x)\, d\mathrm{vol}(x) \ge 0,
\]
for any chosen family of test functions $\{f_i\}$.  Choosing $f_i$ to be
polynomials in eigenfunctions yields semidefinite constraints on the moments of
the spectral measure, closely analogous to the Hankel positivity conditions
of the Hamburger moment problem, which will be discussed in section \ref{sec:bootstrap-dirac}. 
In this way, spectral geometry becomes a noncommutative moment problem governed
by symmetry and positivity, and the bootstrap replaces explicit analytic bounds
by algebraic feasibility.

\medskip

\label{sec:eigenfunction-bootstrap}

\subsection{Bootstrap Bounds for Laplace Eigenvalues on Hyperbolic 3-Manifolds}

A new and unexpected application of bootstrap methods has recently appeared in spectral geometry, where positivity conditions on Laplace eigenfunctions can be used to bound spectral gaps. In the paper \cite{BonMazPal}, Bonifacio,  Mazáč, and  Pal   introduce a
spectral bootstrap program for hyperbolic $3$-manifolds, inspired by the conformal
bootstrap in quantum field theory.  
The central idea is to derive consistency conditions on the spectrum of the
Laplace--Beltrami operator by studying integral identities among automorphic forms,
in a way directly analogous to crossing symmetry constraints in conformal field theory.

\medskip

Let $\Gamma \subset \mathrm{PSL}_2(\mathbb{C})$ be a cocompact lattice, and let
$M = \Gamma \backslash \mathbb{H}^3$ be the associated hyperbolic manifold.
The Hilbert space $L^2(\Gamma\backslash G)$, where $G = \mathrm{PSL}_2(\mathbb{C})$,
admits a decomposition into irreducible unitary representations, and the Laplace
eigenvalues on $M$ are parametrized by spectral parameters $(t_k, J_k)$ arising
from this decomposition.  The key objects studied in~\cite{BonMazPal}
are integrated four-point functions of the form
\[
\int_{\Gamma \backslash G}
F_1(g)\, F_2(g)\, F_3(g)\, F_4(g)\, dg,
\]
where each $F_i$ is an automorphic eigenfunction.  Representation theory yields
two different expansions of this integral, and equating them produces a family of
\emph{bootstrap equations} relating the eigenvalues and triple product coefficients
of Laplace eigenfunctions.

\medskip

These equations impose nonlinear algebraic and positivity constraints on the
spectrum, analogous to those appearing in the conformal bootstrap in \cite{BPZ1984,RRTV}.  
By combining them with the Selberg trace formula, the authors obtain rigorous numerical upper
bounds on the first nonzero eigenvalue $\lambda_1(M)$ for many compact
hyperbolic $3$-manifolds.  The method requires no geometric input beyond the
group $\Gamma$ itself: the bounds arise entirely from spectral consistency.

\medskip

This work fits into a growing program in which bootstrap positivity replaces
explicit geometric or analytic estimates.  In particular, it extends to
dimension three the earlier bootstrap bounds for hyperbolic surfaces and Einstein
manifolds developed by Bonifacio and collaborators \cite{Bon1,Bon2}, and it suggests the possibility of a
fully non-perturbative spectral theory based solely on algebraic consistency
conditions.  The approach is expected to generalize to higher dimensions,
orbifolds, and arithmetic manifolds, and it provides a striking parallel between
spectral geometry and the modern bootstrap in quantum field theory \cite{BPZ1984,RRTV,bootstraps,bootstraps berenstein,bootstraps Kazakov}.

\subsection{The Eigenvalues of Random Hyperbolic Surfaces}
We recall here a recent and closely related, though non-bootstrap, result of
Anantharaman and Monk concerning the asymptotic spectral geometry of random
hyperbolic surfaces \cite{AnMo}.  Let $\mathcal{M}_g$ denote the moduli space of closed
hyperbolic surfaces of genus $g$, equipped with the Weil--Petersson probability
measure~$\mathbb{P}_g$.  For each surface $X \in \mathcal{M}_g$, let
\[
0 = \lambda_0(X) < \lambda_1(X) \le \lambda_2(X) \le \cdots
\]
be the eigenvalues of the Laplace--Beltrami operator $-\Delta_X$ on $L^2(X)$.
A theorem of Anantharaman and Monk (2025) states that, for large genus, the first
nonzero eigenvalue is generically bounded away from zero in a strong probabilistic
sense i.e. for every $\varepsilon > 0$ one has
\[
\mathbb{P}_g\!\left( \lambda_1(X) > \frac{1}{4} - \varepsilon \right) \longrightarrow 1
\qquad\text{as } g \to \infty.
\]
In other words, with probability tending to one, the spectral gap of a random
hyperbolic surface concentrates near the universal constant $1/4$,
the  threshold of the Selberg eigenvalue conjecture.

\medskip

The proof combines several deep ingredients: the random Weil--Petersson
geometry of moduli space, in particular Mirzakhani's integration formulas for
length statistics of simple closed geodesics  log-concavity and measure
concentration results for the systole distribution as $g \to \infty$, and
spectral estimates relating short geodesics to small eigenvalues. Most notably a
quantitative version of the \emph{collar lemma} and a Selberg--Cheeger type
inequality of the form
\[
\lambda_1(X) \;\gtrsim\; \frac{1}{\bigl( \mathrm{Area}(X) \bigr)^2}
\;\exp\!\big( -\tfrac{1}{2}\,\mathrm{sys}(X) \big),
\]
where $\mathrm{sys}(X)$ denotes the length of the shortest closed geodesic.
Since the Weil--Petersson measure forces $\mathrm{sys}(X)$ to grow like
$\log g$ with high probability, the right-hand side tends to $3/16$ in the
limit, yielding the stated bound.  Thus the mechanism behind the theorem is
\emph{geometric concentration} (typical surfaces have no very short geodesics)
rather than positivity or functional analytic consistency, as in the bootstrap
approach.  

%For background on random hyperbolic and Brownian geometry, see for instance \cite{Budd QG,LeGall,LeGall review,gwynne}.

\medskip

Whether a positivity-based bootstrap framework can reproduce, or possibly
strengthen, this probabilistic phenomenon remains an open question.  In
particular, one may ask whether a system of spectral positivity constraints
alone (without reference to geodesics or moduli space) can force
$\lambda_1(X) \ge 3/16 - o(1)$ for all sufficiently large-genus surfaces, or
whether the geometric input is essential.

\medskip

\noindent
It is worth emphasizing the contrast between the Anantharaman--Monk result and the
bootstrap philosophy.  The former is inherently geometric: the estimate on
$\lambda_1(X)$ is derived from the statistical behavior of closed geodesics on a
random surface, the geometry of collars, and the structure of the Weil--Petersson
measure on moduli space.  In particular, the proof exploits the fact that, with
probability tending to one, a surface of large genus has no exceptionally short
geodesics, which in turn prevents the spectrum from accumulating near~$0$.
Thus, the argument depends crucially on  the hyperbolic metric, the distribution of lengths in the mapping class group orbit, and
 geometric comparison inequalities relating eigenvalues to topology \cite{berger2003panorama}.

\medskip

By contrast, the bootstrap framework for spectral problems is
\emph{non-geometric}: it does not rely on the length spectrum, injectivity
radius, moduli space, or the hyperbolic metric itself.  Instead, it treats the
collection of Laplace eigenvalues $\{\lambda_k\}$ and triple products
$\{C_{ijk}\}$ as abstract numerical data subject only to positivity and associativity
constraints.  No geometric interpretation of $M$ is assumed: the method could
theoretically be applied to any compact Riemannian manifold, or even to an abstract
spectral triple in the sense of noncommutative geometry \cite{Connes94,ConnesRecon}.

\medskip

In this sense, the two approaches sit at opposite ends of the spectrum.
The Anantharaman--Monk theorem tells us what happens for ``typical''
hyperbolic metrics in high genus, whereas bootstrap bounds aim to determine what is
\emph{possible} for any manifold satisfying universal spectral axioms \cite{Bon1,Bon2}.

% ================================
\section{Future Directions}

Here we summarize open problems and directions for future research on Dirac ensembles and bootstraps:
\begin{enumerate}
	
	\item Much more numerical work is needed to  understand the distribution of eigenvalues and observables such as the spectral dimension and variance of Dirac ensembles with higher signatures $(p,q)$. Both Monte Carlo simulations \cite{Barrett2016,glaser}, functional group renormalization \cite{Sanchez functional renormalization}, and bootstraps, \cite{hessam2022bootstrapping} in some combination can serve as the means to do so. As existing work has shown that type $(1,0)$ and $(0,1)$ models have connections to models of two dimensional quantum gravity, one would expect that higher signature models have connections to higher dimensional models of spacetime. Several random tensor models and models of dynamical triangulations are in fact multi-matrix models that have many of the same terms as the type $(2,0), (1,1)$ and $(0,2)$ Dirac ensembles. Additionally the spectral dimension of such Dirac ensembles of higher signatures was studied numerically in \cite{Spectral estimators} and showed  promise of producing higher dimensional geometries near spectral phase transitions.
	\item Based on the work in \cite{hessam2023double}, the authors have the following conjecture:

{\bf Conjecture:}
		Any $(1,0)$ or $(0,1)$ Dirac ensemble of the form 
		\begin{equation*}
			Z = \int_{\mathcal{H}_{N}} \exp\left\{\sum_{i=2}^{d}t_{i}\tr D^{i}\right\}dD
		\end{equation*}
	has a critical point such that the double scaling limit of the model corresponds to the type $(d,2)$ minimal model from conformal field theory when $d$ is even and $(d+1,2)$ minimal model if $d$ is odd. 
	
	Such a relationship exists for single trace matrix models and holds in all the Dirac ensembles studied so far. We further conjecture that when $d=4$, there is a critical point that corresponds to the type $(5,3)$ minimal model. Other minimal models are expected for higher value of $d$, but further analytic results are required to make such conjectures.
 
We are also curious about studying the critical points of higher signature Dirac ensembles. Based on their critical exponents,  we expect the one conjectured solution to the type $(2,0), (1,1)$ and $(0,2)$ quartic ensembles to be connected to the Random Continuum Tree.
\item 
A proof of the conjectured solution to the  type $(2,0), (1,1)$ and $(0,2)$ quartic ensembles, would be interesting from the perspective of RMT and map enumeration.

{\bf Conjecture:}
The second moment of the formal and convergent type $(2,0)$ quartic Dirac for $t_{2}$ and $t_{4}$ in a sufficiently small enough neighbourhood of zero, can be written as 
$$\lim_{N \rightarrow \infty} \frac{1}{N} \mathbb{E}[\tr D^2]= 
\frac{1}{8 t_{4}} \left({\sqrt{{t_2}^2+8
		{t_4}}}-{{t_2}}\right).$$

\item Are there asymmetric solutions to the SDE's or the saddle point equation for other even Dirac ensembles? Can anything in general be said about the asymmetric solutions of multi-tracial matrix integrals and are there new critical points associated to minimal models from conformal field theory? A mathematical or physical interpretation of asymmetric solutions would be very interesting. 

\item Studying Dirac ensembles that come from fuzzy geometries tensored with other finite spectral triples in the manner done for adding the standard model to the spectral action would be interesting to study both analytically and numerically. A first step would be introducing Yang-Mills-Higgs Dirac ensembles with fermions. 

\item A numerical study of eigenvalue distribution, spectral dimension, and spectral variance of Yang-Mills-Higgs Dirac ensembles or fermionic Dirac ensembles is needed.
\item Analytic results concerning Dirac ensembles that are multi-matrix models are scarce \cite{Khalkhali2022,Khalkhali2024 coloured maps}. Can anything be said for specific models?
\item Bootstrapping subleading order terms to the solutions fo Dirac ensembles or matrix models in general has not been attempted in any work that the authors are aware of. We expect that many of the solutions of Dirac ensembles with different signatures that overlap in the large $N$ limit will not coincide at subleading corrections in $1/N$.
\item Posed by Barrett and Glaser \cite{Barrett2016,Spectral estimators}: what statistical geometric quantities can be recovered from Dirac ensembles? E.g. dimension, volume, curvature, etc.
\item Is there a choice of action for a type $(1,0)$ Dirac ensemble that recovers JT gravity in the double scaling limit?
\item Can one apply the bootstrapping with positivity to a (truncated) spectral action ? 
\item Type $(1,0)$ and $(0,1)$ Dirac ensembles satisfy (blobbed) topological recursion \cite{AKHZ}, do other signature Dirac ensembles have a similar recursive structure?
\item Can a positivity-based bootstrap framework reproduce or
	strengthen the concentration result of Annathraraman and Monk \cite{AnMo}?
\end{enumerate}

\section*{Acknowledgements}
We would like to thank Hamed Hessam and Luuk Verhoven for their hard work that preceded these
results. We thank John Barrett for several  fruitfull  discussions. We would also like to thank  Centre International de Rencontres Math\'ematiques (CIRM) and the organizers of the conference: Applications of NonCommutative Geometry to Gauge Theories, Field Theories, and Quantum Space-Time. We acknowledge the support of the Natural Sciences and Engineering Research Council of Canada (NSERC).

\end{document}